\documentclass[twocolumn,bibyear]{aa}

\usepackage{graphicx}
\usepackage[varg]{txfonts}
\usepackage{natbib}
\usepackage[colorlinks=true,     linkcolor=blue, citecolor=blue, filecolor=blue, urlcolor=blue]{hyperref}
\usepackage{epsfig}
\usepackage{color}
\usepackage{xcolor}

\def\cc{ç}

\newcommand   {\about} {\mbox{$\sim$}}
\newcommand   {\mic}   {\mbox{$\mu$m}}
\newcommand   {\tim}   {\mbox{$\times$}}
\newcommand   {\hh}    {\mbox{H$_2$}}

\renewcommand {\deg}   {\mbox{$^\circ$}}
\newcommand   {\arcm}  {\mbox{$^\prime$}}
\newcommand   {\arcs}  {\mbox{$^{\prime\prime}$}}
\newcommand   {\pccm}  {\mbox{cm$^{-3}$}}
\newcommand   {\pscm}  {\mbox{cm$^{-2}$}}
\newcommand   {\kms}   {\mbox{km\,s$^{-1}$}}

\renewcommand {\ga}    {\mbox{\rlap{\hbox{\lower5pt\hbox{$\sim$}}}\hbox{$>$}}}
\renewcommand {\la}    {\mbox{\rlap{\hbox{\lower5pt\hbox{$\sim$}}}\hbox{$<$}}}

\newcommand{\cmark}{\color{black}}
\newcommand{\umark}{\color{black}}

\begin{document}

\title{An 18 -- 25 GHz spectroscopic survey of dense cores in the
  Chamaeleon~I molecular cloud} 

\author{Dariusz~C.~Lis\inst{1}, William~D.~Langer\inst{1},
  Jorge~L.~Pineda\inst{1}, Kahaan~Gandhi\inst{2,1},
  Karen~Willacy\inst{1}, Paul~F.~Goldsmith\inst{1},
  Susanna~Widicus~Weaver\inst{3,4}, Liton~Majumdar\inst{5,6},
  Youngmin~Seo\inst{7}, Shinji~Horiuchi\inst{8}, Cheikh~Bop\inst{9,10},
  and Fran\cc ois~Lique\inst{9}}

\institute{Jet Propulsion Laboratory, California Institute of
  Technology, 4800 Oak Drove Drive, Pasadena, CA 91109, USA
  \and Haverford College, 370 Lancaster Avenue, Haverford, PA 19041, USA 
  \and Department of Astronomy, University of Wisconsin-Madison, 475 N
  Charter St, Madison, WI 53706, USA
  \and Department of Chemistry, University of Wisconsin-Madison, 1101
  University Ave, Madison, WI 53706, USA
  \and Exoplanets and Planetary Formation Group, School of Earth and
  Planetary Sciences, National Institute of Science Education and
  Research, Jatni 752050, Odisha, India 
  \and Homi Bhabha National Institute, Training School Complex,
  Anushaktinagar, Mumbai 400094, India 
  \and Aerospace Corporation, 4745 Lee Road, Chantilly, VA 20151, USA 
  \and CSIRO Space \& Astronomy/NASA Canberra Deep Space Communication
  Complex, PO Box 1035, Tuggeranng ACT 2901, Australia
  \and Univ. Rennes, CNRS, IPR (Institut de Physique de Rennes) – UMR
  6251, 35000 Rennes, France
  \and Nantes Université, CNRS, CEISAM, UMR 6230, F-44000 Nantes,
  France}

\date{Received 24 January 2025; accepted 3 Mar 2025}

\abstract{ The presence of over 300 molecules in the interstellar
  medium, the majority of them organic compounds, raises the question
  of the extent to which protostellar chemistry is responsible for
  organic molecules in solar system bodies (e.g., comets, asteroids,
  planets). The majority of systematic surveys for organic molecules
  in cold cores have focused on the TMC-1 core in the Taurus complex,
  along with lesser surveys of other protostellar cores in the
  northern hemisphere facilitated by the presence of several
  telescopes available for surveys below 45 GHz, where most organic
  molecules have relatively strong emission under conditions in cold
  cores. A few southern hemisphere sources have been surveyed at
  wavelengths between 7 and 1 mm. Here we extend the survey for
  organics in the southern hemisphere to 1.3 cm by observing two cores
  in the Chamaeleon complex using NASA's Deep Space Network 70-m
  antenna in Canberra, Australia, over the frequency range of 18 to 25
  GHz. In the Chamaeleon complex we surveyed the class 0 protostar
  Cha-MMS1 and the prestellar core Cha-C2, which represent two stages
  in the evolution of dense cores. We detect several molecules
  including HC$_3$N, HC$_5$N, C$_4$H, CCS, C$_3$S, NH$_3$, and
  c-C$_3$H$_2$. A longer cyanopolyyne, HC$_7$N, is detected with high
  confidence via spectral stacking analysis. While molecular column
  densities in the two Chamaeleon cores are typically an order of
  magnitude lower compared to the cynaopolyyne peak in TMC-1, the
  molecular abundance ratios are in general agreement with the TMC-1
  values. The two exceptions are c-C$_3$H$_2$, which is enhanced by a
  factor of \about 25 with respect to cyanopolyynes in the Chamaeleon
  cores, and ammonia, which is enhanced by a factor of \about 125. The
  deuterated species c-C$_3$HD is detected in both cores, with a high
  D/H ratio of $\about 0.23$ in c-C$_3$H$_2$. A rare isotopologue of
  ammonia, $^{15}$NH$_3$, is also detected in Cha-MMS1 suggesting a
  high $^{14}$N/$^{15}$N ratio of $\about 690$ in ammonia. However,
  this ratio may be artificially enhanced due to the high optical
  depth of the $^{14}$NH$_3$ (1,1) line, which increases the effective
  source size. We use the detections of ammonia, cyanopolyynes, and
  far-infrared dust continuum to characterize the density and
  temperature in the Chamaeleon cores and calculate the molecular
  column densities and their relative ratios. The ring molecule
  benzonitrile, a tracer for the non-polar molecule benzene, is not
  detected in either Chamaeleon core. The $3 \sigma$ upper limits for
  the benzonitrile column density achieved are a factor of 2 higher
  than the value derived for TMC-1 and the corresponding upper limits
  for the relative abundance of benzonitrile with respect to HC$_5$N
  are a factor of 3 higher than the TMC-1 value. }


\titlerunning{Dense cores in Chamaeleon}
\authorrunning{Lis et al.}

\maketitle
\nolinenumbers

\section{Introduction}
\label{sec:intro}

An extrasolar origin has been proposed as a possible source of complex
organic molecules and water detected in primitive solar system bodies
(e.g., \citealt{hanni22}). Many of the same molecules have been
detected in the interstellar medium, in particular in protostellar
cores and protoplanetary disks. However, the chemical pathways for
their production via gas phase and grain catalysis from atomic and
molecular species are not well established, nor are the mechanisms for
their incorporation into solar bodies such as comets and planets. To
answer these questions, searches have been conducted to detect
molecules in both the gas phase and in ices in a variety of
interstellar sources. To date, astronomical observations have detected
over 320 individual molecular species in the gas phase in
interstellar, protostellar, and circumstellar environments
\citep{mcguire2022}.\footnote{See also http://astrochymist.org}
Furthermore, the majority of these detections are of organic
molecules, made of carbon bonded with other elements or
other carbon atoms. Most of these discoveries have been detections
enabled by radio astronomy of spectral signatures of molecules ranging
in size from two (e.g., OH and CO) to thirteen atoms in the long chain
HC$_{11}$N and recently detected ring molecule benzonitrile,
C$_6$H$_5$CN \citep{mcguire2018}, and even larger
polycyclic aromatic hydrocarbons C$_{10}$H$_7$CN \citep{mcguire2021},
C$_{12}$H$_{8}$ \citep{cernicharo24}, or C$_{16}$H$_9$CN
\citep{wenzel24}. 

Systematic searches for species in a broad range of sources is needed
so that a complete picture of their abundances can be developed in
order to determine the chemical pathways forming complex organic
molecules. In the northern hemisphere the "Astrochemical Surveys at
IRAM" (ASAI) carried out an unbiased spectral survey between 80 and
272 GHz of 10 sources spanning a range of evolutionary states,
including the \cmark starless \umark core TMC-1 in the Taurus
Molecular Cloud \citep{lefloch2018}. TMC-1 is one of the richest
interstellar sources of organic molecules and it has also been the
target of dedicated radio observations of emission spectra at longer
wavelengths. The two current systematic studies of TMC-1 at longer
wavelengths are QUIJOTE, Q-Band Ultrasensitive Inspection Journey to
the Obscure TMC-1 Environment, using the Yebes telescope
\citep{cernicharo2021} and GOTHAM, GBT Observations of TMC-1: Hunting
Aromatic Molecules, at the Green Bank Telescope \citep{mcguire2018}.
Most of the other sources that have been systematically surveyed are
highly evolved regions of massive star formation, such as Sgr B2 and
Orion, or evolved stars such as IRC+10216. However, broad spectral
surveys of cold protostellar cores, even those residing in regions of
massive star formation, such as is thought to be the environment in
which the solar system formed, are much less explored. The excitation
conditions at low temperatures and high densities in these cold cores
favor observational surveys at wavelengths $\gtrsim$ 3mm, and for the
heavier organics $\gtrsim$ 7~mm.

In the southern hemisphere far fewer sources have been studied
systematically at long wavelengths. The Australia Telescope Compact
Array has conducted a Q-band survey of Sagittarius B2 and detected
over 53 molecular species \citep{corby2015}. However, Sgr B2 is an
atypical Galactic source, and so its chemistry may be more
characteristic of energetic massive star forming cores rather than the
pathway to solar-type planet forming disks. In the southern hemisphere
the best studied low-mass protostar at long wavelengths is Chameleon
MMS1 which was observed by \cite{kontinen00} at 3~mm wavelength with
the Swedish-ESO Submillimeter Telescope (SEST) and by
\cite{cordiner2012} at 7 mm wavelength with the ATNF Mopra 22m
telescope. The Chameleon complex is also a target of the JWST "Ice
Age" ERS program \citep{mcclure23} ultimately making it possible to
compare gas phase and ice phase chemistry.

\cmark Located at a distance of 190 pc \citep{galli21}, \umark
Chamaeleon I is one of the closest low-mass star forming regions in
the southern hemisphere (\citealt{belloche11} and references therein).
Embedded within the Chamaeleon I ridge traced by millimeter dust
continuum emission are a known prestellar core, a Class 0 protostar,
\cmark a binary protostellar system Ced 110 IRS4 \citep{rocha25}. and
an edge-on Class II protoplanetary disk HH 48 NE
\citep{sturm24}. \umark To cover a range of physical environments most
relevant for the complex organic molecule (COM) chemistry, we
selected the Class 0 protostar (Cha-MMS1) and the prestellar core
(Cha-C2) as targets of our observations.

In this paper we extend the survey of Chameleon to radio K-band using
NASA's Deep Space Network (DSN) 70-m antenna in Canberra, Australia
(DSS-43) covering the 18 to 25 GHz frequency range. \cmark The two
Chamaeleon sources studied here have different ages, with the class 0
protostar Cha-MMS1 being more evolved than the prestellar core Cha-C2.
Moreover, being surrounded by a group of YSO's, the two cores are
located in a different environment than the best studied starless
core, TMC-1. The observations thus provide interesting insights into
the effects of environment and age on the dense core chemistry. \umark
We use multi-wavelength \emph{Herschel} observations of the dust
continuum emission and multiple transitions of ammonia, HC$_3$N, and
HC$_5$N to derive the kinetic temperatures and densities of the two
cores.

In Section~\ref{sec:dsn} we discuss the DSS-43 observations and in
Section~\ref{sec:properties} we derive the properties of the cores and
molecular column densities. In Section~\ref{sec:stacking} we use a
Markov Chain Monte Carlo (MCMC) analysis to identify features in the
spectral scans that are too weak to identify from individual spectral
lines.
In Section~\ref{sec:discussion} we discuss the differences between the
Chamaeleon sources and TMC-1. In Section~\ref{sec:conclusion} we
summarize our results and describe the next steps needed to trace the
evolution of prebiotic compounds in protostellar cores.

\begin{figure}
\centering

\includegraphics[trim=2.5cm 2.5cm 3cm 5cm, clip=true, width=\columnwidth,angle=0]{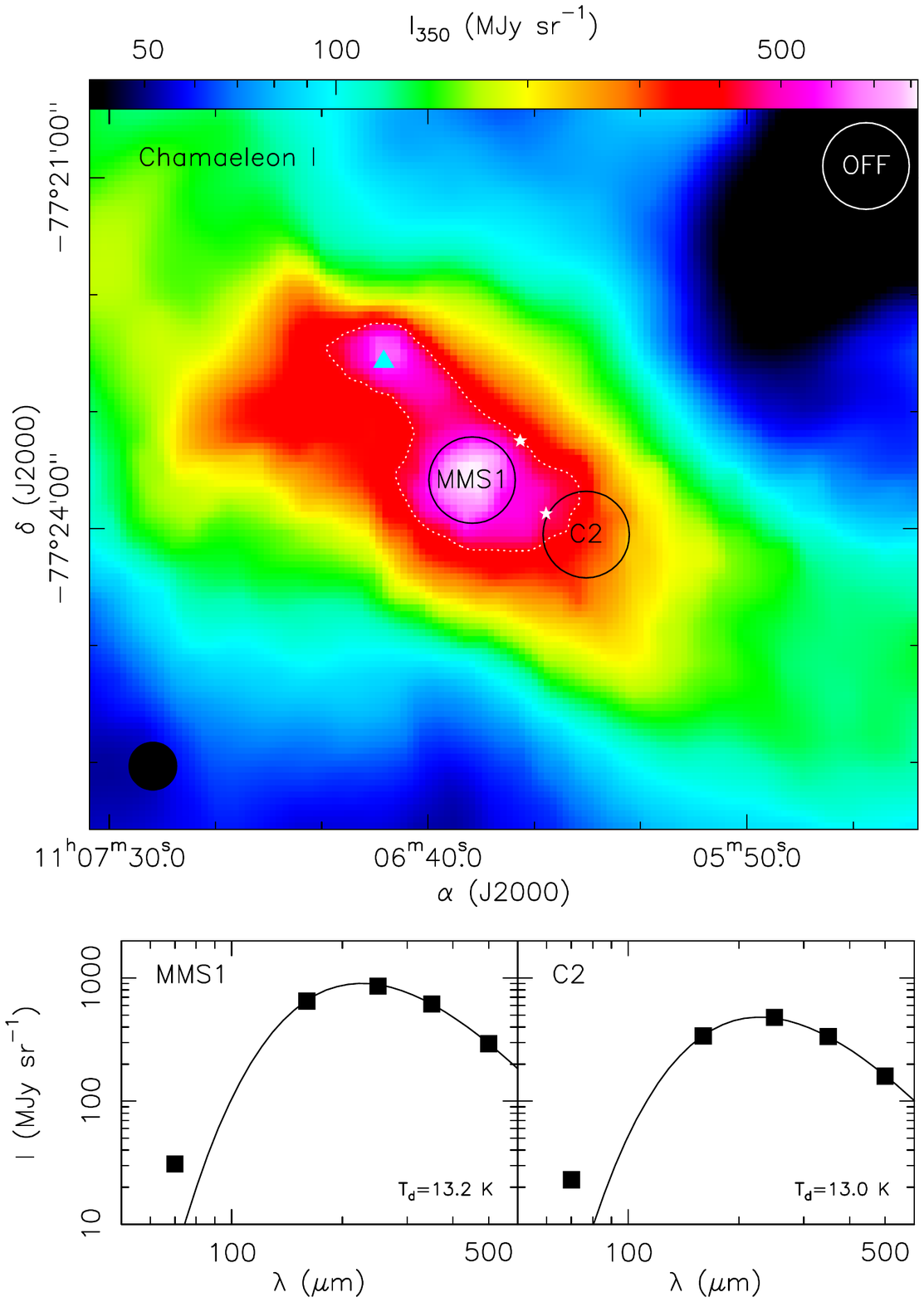} 

\caption{\emph{Herschel}/SPIRE image of 350~$\mu$m dust continuum
  emission toward the central part of the Chamaeleon I cloud. The
  dotted contour outlines the extent of the dust emission at 50\% of
  the peak. Black circles mark the Cha-MMS1 and Cha-C2 pointings and
  the white circle indicates the reference position used for the
  DSN 
  observations. The size of the two circles corresponds to the FWHM
  beam size of the Canberra telescope (45$^{\prime\prime}$). The black
  circle in the lower-left corner shows the FWHM size of the SPIRE
  beam (25.2$^{\prime\prime}$). White asterisks mark locations of
  NIR38 and J110621228, two background stars with ice spectra studied
  by \cite{mcclure23}, and the cyan triangle marks the location of the
  binary protostar Ced 110 IRS4 (\citealt{rocha25}). The lower panels
  show the dust continuum SED in the Canberra beam, based on PACS and
  SPIRE observations. The black curves are modified blackbody fits to
  the SPIRE and PACS 160~$\mu$m \cmark surface brightness, \umark as
  described in the text. Typical flux calibration uncertainties are
  5\%.}
\label{fig:spire}
\end{figure}

\section{Deep Space Network observations}
\label{sec:dsn}

The Deep Space Network radio telescopes had a nearly decade-long
history of contributions to the search for organic molecules in the
northern hemisphere and study of their formation environment beginning
in the early 1990s \citep{langer95, velu95, kuiper96, langer97,
  langer98, peng98, dickens01}. The DSN technical capabilities for
astrochemistry research have improved significantly in recent years
with the installation of a new cryogenic dual-horn dual-polarization
17–27 GHz receiver at the Deep Space Station 43 (DSS-43) in Canberra,
Australia \citep{kuiper19}, with its broadband digital spectrometer
\citep{virkler20}. The wide instantaneous bandwidth allows
observations of multiple transitions of heavy species which can be
used to characterize the density and temperature of the gas using
radiative transfer models, and dynamical information from the line
width and velocity at the line peak. The spectrometer provides
sufficiently high spectral resolution to resolve molecular line shapes
even in the coldest regions ($\sim 10$~K) of protostellar cores.

Several observing runs of the Chamaeleon~I cloud using the DSN 70-m
antenna at Canberra were carried out in February -- September, 2022.
Figure~\ref{fig:spire} shows the overall morphology of the region as
traced by 350~\mic\ dust emission observed with the SPIRE instrument
on \emph{Herschel}. The black circles show the two DSS-43 pointings,
centered on the class 0 protostar MMS1 and the prestellar core C2. The
J2000 source coordinates for Cha-MMS1 are $\alpha$ =
$11^{\rm h}06^{\rm m} 33.13^{\rm s}$, $\delta$ =
$-77\deg 23^\prime 35.1^{\prime\prime}$ and for Cha-C2 $\alpha$ =
$11^{\rm h}06^{\rm m} 15.51^{\rm s}$, $\delta$ =
$-77\deg 24^\prime 04.9^{\prime\prime}$. All observations were carried
out in a position-switching mode, using a reference position
~5\arcm\ N-W of Cha-MMS1, in a direction perpendicular to the extent
of the dust ridge (at $11^{\rm h} 05^{\rm m} 30.00^{\rm s}$,
$-77\deg 21^{\prime} 00.0^{\prime\prime}$; white circle in the
upper-right corner of Fig.~\ref{fig:spire}). Circle sizes correspond
to the FWHM beam size of the Canberra telescope at 22 GHz, 45\arcs.
The filled black circle in the lower-left corner shows the SPIRE
350~\mic\ FWHM beam size of 25.2\arcs.

The 350~\mic\ continuum emission in Chamaeleon~I is extended on scales
comparable to the DSS-43 beam size (see 50\% white dotted contour in
Fig.~\ref{fig:spire}). \cite{belloche11} derive FWHM source sizes of
$55\arcsec \times 49\arcsec$ and $98\arcsec \times 46\arcs$ for
Cha-MMS1 and Cha-C2, respectively, based on observations of 870~$\mu$m
dust continuum emission using the APEX telescope. We thus assume
average source sizes of 52 and 67$^{\prime\prime}$ in the column
density calculations.

The DSN Canberra K-band digital spectrometer processes sixteen 1-GHz
wide bands, split into 8 separate bands from 18 to 26 GHz for each
polarization \citep{virkler20}. Each band consists of 32,768 channels
with a 30.5 kHz resolution, corresponding to a velocity resolution of
$0.35 - 0.49$~km\,s$^{-1}$, depending on the frequency. A typical FWHM
line width in Cha-MMS1 is $0.6 - 0.9$~km\,s$^{-1}$, depending on the
species. All lines are thus spectrally resolved. A typical system
temperature at the elevation of the source was 77 K. The total
on-source integration time was about 14 hours per source. For the
seven lowest-frequency bands, 2 instrumental polarizations were
observed, doubling the effective observing time. At frequencies above
25~GHz, only one instrumental polarization was available. The
resulting spectra thus have higher noise and are not included in the
analysis.

The raw data from the spectrometer were processed into calibrated
ON-OFF spectra using the standard DSS-43 data reduction pipeline.
\cmark The system temperature was continuously monitored using a power
meter, scaling with a factor derived using a noise diode and an
ambient load before the observation; it is used in the standard ON-OFF
calibration to obtain spectra in the antenna temperature units,
$T_A^*$. We refer the reader to \cite{kuiper19} for details on the
absolute system and receiver temperature calibration. A relative gain
correction was applied to the data to account for antenna deformations
as a function of elevation. The gain dependence on elevation was
determined using measurements of a flux calibrator at different
elevations, showing a peak at about 45$^{\circ}$ elevation.
\cite{kuiper19} fitted a third order polynomial to the data, but we
used a second order fit, as discussed in the ATNF Tidbinbilla 70-m
Radio Telescope
Guide\footnote{https://www.atnf.csiro.au/resources/observing/observers/tidbinbilla/
  tid\_obs\_guide/}, which provides more accurate values for
observations taken at high elevations. \umark

The beam efficiency was not measured directly during our observations.
To convert the observed spectra to the main beam brightness
temperature units, we use the main beam efficiency of $\eta_{mb}=50$\%
\citep{pineda19} rather than the measured DSN aperture efficiency of
$\eta_A = 35.5$\%. This choice is justified given the expected extent
of the molecular emission. Moreover, the absolute intensity calibration is
not critical, as opacity effects are small or moderate for all
detected lines except for ammonia, and we use only the relative
abundance ratios among the molecules rather than absolute abundances
with respect to H$_2$ in the comparison with TMC-1.

Subsequent data reduction was carried out using the IRAM CLASS data
reduction software\footnote{https://www.iram.fr/IRAMFR/GILDAS/}.  The
data reduction included blanking of noisy channels near the band edges
and removing 3rd order polynomial baselines from individual scans,
fitted across the full frequency range of each band. The resulting
baseline-removed spectra were then averaged with 1/$\sigma^2$ weighting
to produce the final spectra used in the analysis.

\begin{figure*}
  \centering
  \includegraphics[trim=2.5cm 3.0cm 6.5cm 2.5cm, clip=true, width=0.9\textwidth]{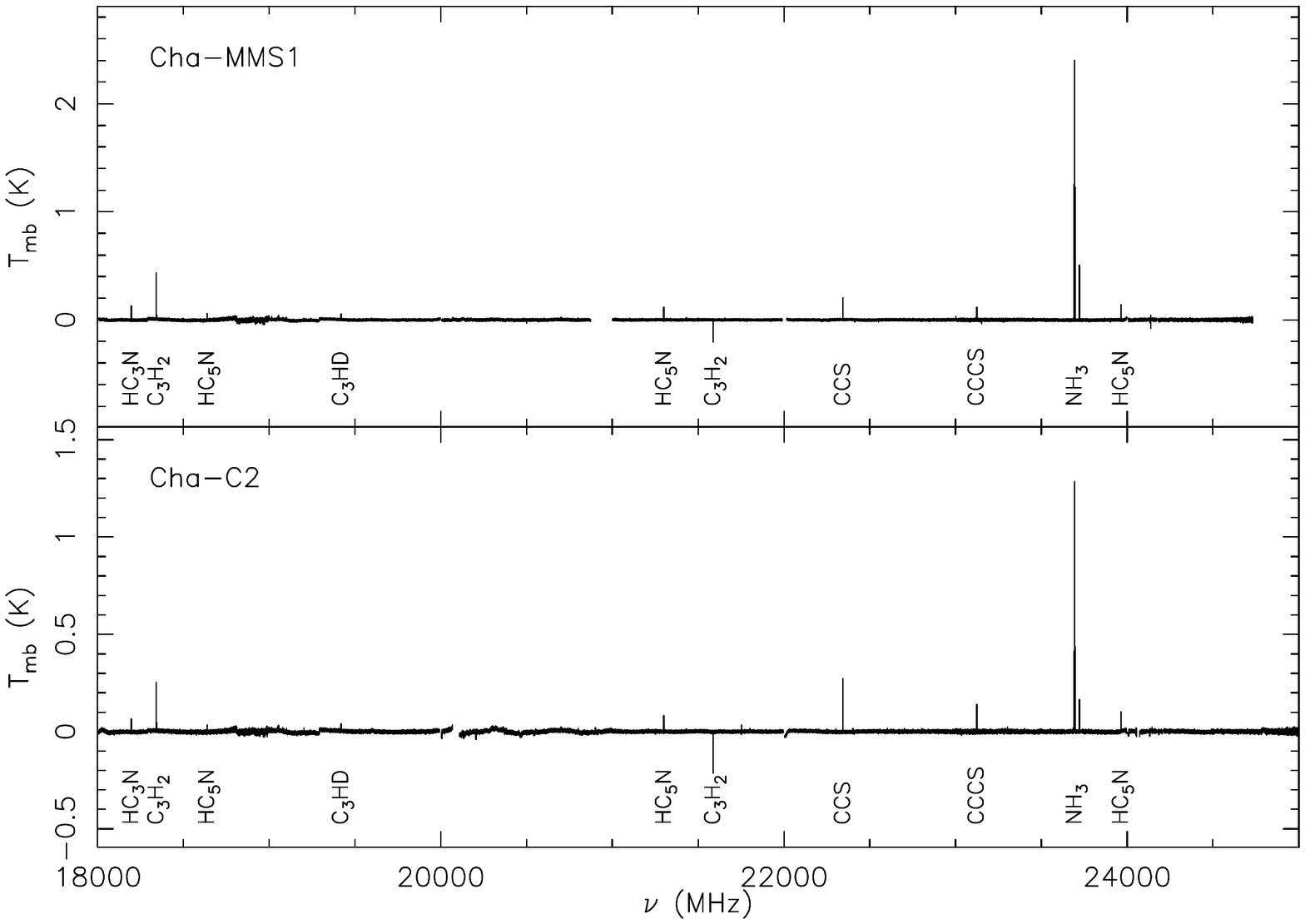}
  \caption{DSN spectra of Cha MMS1 and Cha C2 (upper and lower
    panel, respectively) corrected for the main beam efficiency.
    Channels with excess noise near subband edges and spurious signals
    have been blanked, resulting in some gaps in the frequency
    coverage. Detected spectral lines are identified (see Tables
    \ref{tab:mms1} and \ref{tab:c2}).} \label{fig:spectra}
\end{figure*}

The resulting full-band spectra of Cha-MMS1 and C2 (in antenna
temperature $T_{mb}^*$ units) are shown in Figure~\ref{fig:spectra}, with
the strongest molecular lines labeled. The noise level is not
uniform, and often increases significantly toward the band edges. In
addition, some noise spikes (“spurs”) are present in the spectra.
Channels with excess noise near band edges and narrow spurious signals
have been blanked, resulting in some gaps in the frequency coverage.
The origin of these artifacts was not investigated, as they do not
affect any of the spectral lines discussed below.

Since in some cases the rms can vary significantly across the subband,
spectra of all lines identified in the broadband survey were
subsequently re-reduced by computing the local noise in the immediate
vicinity of each line and using this value in the weighted average.

\begin{table*}
\begin{center}  
\caption{Parameters of fits to lines observed in Chamaeleon MMS1.} 
\label{tab:mms1}
\begin{tabular}{ccccccc}
\hline \hline 
  \rule[-3mm]{0mm}{8mm} Transition & $\nu_o$ & $t$ & $\sigma$
  & $\int T_{mb} dv$ & $V_{LSR}$ & $\Delta V$ \\
  & (MHz) & (hr) & (mK) & (mK kms$^{-1}$) & (kms$^{-1}$) & (kms$^{-1}$)\\
  \hline 
  HC$_3$N $(2-1)$ & 18196.2300 & 26.7  
    & 3.8 & 338. (10.) & 3.96 (0.01) & 0.84 (0.01) \\
  c-C$_3$H$_2$ $(1_{10}-1_{01})$ & 18343.1430 & 26.7 
    & 3.9 & 513. (4.0) & 3.83 (0.01) & 0.84 (0.01) \\
  HC$_5$N $(7-6)$ & 18638.6164 & 26.7
    & 4.5 & 86.7 (4.5) & 3.92 (0.03) & 0.82 (0.02) \\
  C$_4$H ($2-1, 5/2-3/2, 2-1)$ & 19014.7204 & 26.7
    & 3.8 & 20.8 (4.5) & 4.18 (0.08) & 0.81 (0.25) \\
  C$_4$H ($2-1, 5/2-3/2, 3-2)$ & 19015.1435 & 26.7 
    & 3.6 & 26.0 (3.7) & 4.15 (0.04) & 0.64 (0.29) \\
  C$_4$H ($2-1, 3/2-1/2, 2-1)$ & 19054.4762 & 26.7
    & 3.8 & 27.5 (6.2) & 4.15 (0.09) & 0.71 (0.19) \\
  c-C$_3$HD $(1_{10}-1_{01})$ & 19418.7272 & 26.7
    & 3.3 & 71.5 (4.5) & 4.12 (0.04) & 1.44 (0.12) \\
  HC$_5$N $(8-7)$ & 21301.2614 & 26.8
    & 3.6 & 115. (6.3) & 4.04 (0.01) & 0.77 (0.03) \\
  c-C$_3$H$_2$ $(2_{20}-2_{11})$ & 21587.4010 & 26.8
    & 2.9 & -199. (2.7) & 4.03 (0.01) &  0.79 (0.01) \\
  CCS $(1-2, 2-1)$ & 22344.0308 & 26.8 
    & 3.0 & 191. (2.6) & 3.88 (0.01) & 0.74 (0.01) \\
  $^{15}$NH$_3$ (1,1) & 22624.9295 & 26.8
    & 3.2 & 16.8 (3.3) & 3.88 (0.11) & 1.04 (0.26) \\
  C$_3$S $(4-3)$ & 23122.9836 & 14.0
    & 4.8 & 88.4 (3.9) & 4.33 (0.02) & 0.73 (0.04) \\
  NH$_3$ (1,1) & 23694.4955 & 14.0
    & 3.5 & 12,690 (47.) & 4.33 (0.01) & 0.64 (0.01) \\
  HC$_5$N $(9-8)$ & 23963.9007 & 14.0
    & 4.3 & 131. (3.8) & 4.30 (0.01) &  0.64 (0.40) \\
  NH$_3$ (2,2) & 23722.6333 & 14.0
    & 4.5 & 648. (4.5) & 4.31 (0.01) & 0.75 (0.01) \\
  NH$_3$ (3,3) & 23870.1292 & 14.0
    & 3.8 & 10.1 (2.2) & 4.26 (0.04) & 0.38 (0.65) \\
\hline
\end{tabular}
\end{center}

Note: Entries in the table are: molecular transition, rest frequency,
total observing time including the two polarizations, final rms in a
single spectrometer channel in the vicinity of the line, integrated
line intensity (opacity corrected for the HFS fits; see Sect.~2.5.3 of the
CLASS user manual,
https://www.iram.fr/IRAMFR/GILDAS/doc/pdf/class.pdf), line velocity,
and line width,. Integrated line intensities have been corrected for
the beam efficiency, but not for the coupling to the source. Values in
parenthesis are 1$\sigma$ fit uncertainties. The opacity of the main
NH$_3$ (1,1) hyperfine component implied by the HFS fit is
$7.63 \pm 0.05$.

\end{table*}

\begin{table*}
\begin{center}  
\caption{Parameters of fits to lines observed in Chamaeleon C2.} 
\label{tab:c2}
\begin{tabular}{ccccccc}
\hline \hline 
  \rule[-3mm]{0mm}{8mm} Transition & $\nu_o$ & $t$ & $\sigma$
  & $\int T_{mb} dv$ & $V_{LSR}$ & $\Delta V$ \\
  & (MHz) & (hr) & (mK) & (mK kms$^{-1}$) & (kms$^{-1}$) & (kms$^{-1}$)\\
  \hline 
  HC$_3$N $(2-1)$ & 18196.2300 & 22.0 
     & 3.4 & 209. (12.) & 3.86 (0.01) & 0.89 (0.04) \\
  c-C$_3$H$_2$ $(1_{10}-1_{01})$ & 18343.1430 & 22.0
     & 3.2 & 344. (3.5) & 3.82 (0.01) & 0.92 (0.01) \\
  HC$_5$N $(7-6)$ & 18638.6164 & 22.0
     & 3.9 & 56.2 (4.0) & 3.83 (0.04) & 0.82 (0.01) \\
  C$_4$H ($2-1, 5/2-3/2, 2-1)$ & 19014.7204 & 22.0
     & 3.7 & 15.9 (4.1) & 4.23 (0.11) & 0.91 (0.24) \\
  C$_4$H ($2-1, 5/2-3/2, 3-2)$ & 19015.1435 & 22.0
     & 3.7 & 23.0 (4.3) & 4.20 (0.10) & 1.08 (0.24) \\
  C$_4$H ($2-1, 3/2-1/2, 2-1)$ & 19054.4762 & 22.0
     & 3.7 & 40.5 (7.8) & 3.93 (0.16) & 1.60 (0.35) \\
  c-C$_3$HD $(1_{10}-1_{01})$ & 19418.7272 & 22.0 
     & 3.5 & 48.7 (5.0) & 4.11 (0.06) & 1.42 (0.21) \\
  HC$_5$N $(8-7)$ & 21301.2614 & 22.0
     & 3.4 & 92.9 (4.7) & 3.95 (0.01) & 0.79 (0.04) \\
  c-C$_3$H$_2$ $(2_{20}-2_{11})$ & 21587.4010 & 22.0 
     & 3.2 & -210. (3.0) & 4.04 (0.01) &  0.88 (0.02) \\
  CCS $(1-2, 2-1)$ & 22344.0308 & 22.0
     & 2.9 & 252. (2.7) & 3.85 (0.01) & 0.81 (0.01) \\
  $^{15}$NH$_3$ (1,1) & 22624.9295 & 22.0
     & 3.3 & $< 11.0^{\rm c}$ \\
  C$_3$S $(4-3)$ & 23122.9836 & 18.5
     & 3.9 & 119. (3.4) & 4.27 (0.01) & 0.78 (0.03) \\
  NH$_3$ (1,1) & 23694.4955 & 18.5  
     & 3.1 & 3,751 (29.) & 4.22 (0.01) & 0.69 (0.01) \\
  HC$_5$N $(9-8)$ & 23963.9007 & 18.5
     & 3.3 & 95.1 (3.0) & 4.23 (0.01) &  0.73 (0.09) \\
  NH$_3$ (2,2) & 23722.6333 & 18.5
     & 3.4 & 192. (30.) & 4.23 (0.01) & 0.70 (0.01) \\
  NH$_3$ (3,3) & 23870.1292 & 18.5
     & 3.5 & $< 11.1^{\rm c}$ \\
  \hline 
  U22902 & 22902.037 & 22.0  
     & 3.6 & 26.9 (3.6) & 4.16 (0.06) & 0.098 (0.17) \\
\hline
\end{tabular}
\end{center}

Note: Entries in the table are: molecular transition, rest frequency,
total observing time including the two polarizations, final rms in a
single spectrometer channel in the vicinity fo the line, integrated
line intensity (opacity corrected for the HFS fits; see Sect.~2.5.3 of the
CLASS user manual,
https://www.iram.fr/IRAMFR/GILDAS/doc/pdf/class.pdf), line velocity,
and line width. Integrated line intensities have been corrected for
the beam efficiency, but not for the coupling to the source. Values in
parenthesis are 1$\sigma$ fit uncertainties. $^{\rm a}$Optically thin
line. $^{\rm b}$Gaussian fit. $^{\rm c}$ 3$\sigma$ upper limit,
integrated over 2.5 -- 5.5 km\,s$^{-1}$ velocity range. An
unidentified $\sim 7.5 \sigma$ feature is detected at 22.902~GHz,
which is not present in the Cha-MMS1 spectrum. The opacity of the main
NH$_3$ (1,1) hyperfine component implied by the HFS fit is
$2.85 \pm 0.04$.

\end{table*}

\begin{figure*}
\centering
\includegraphics[trim=2.5cm 3.0cm 4.5cm 1.5cm, clip=true, width=0.8\textwidth,angle=0]{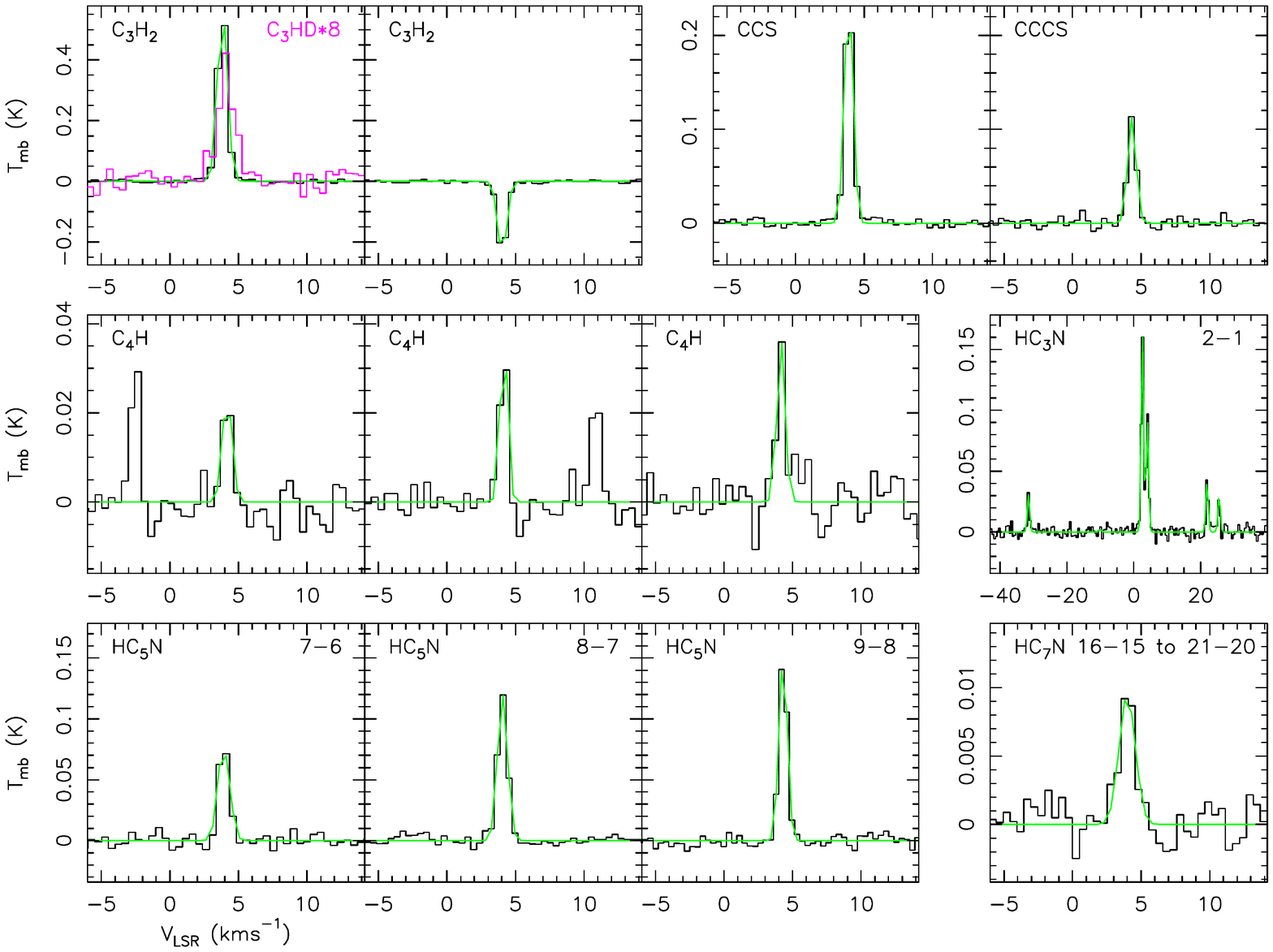} 
\caption{Spectra of molecular lines other than ammonia detected in
  Chamaeleon MMS1 (black and magenta histograms) with fits 
  shown in green.}
\label{fig:spemms1}
\end{figure*}

\section{Results}
\label{sec:properties}

In this section we discuss the physical conditions in the Chamaeleon
cores based on prior observations. We then derive LTE, and for some
species non-LTE, molecular column
densities and molecular abundance ratios and compare them with those
toward the cyanopolyyne peak in TMC-1 \citep{gratier16}. TMC-1, a
target of the QUIJOTE and GOTHAM surveys \citep{cernicharo2021,
  mcguire2018}, is the best studied dense core and a reference source for
astrochemical studies. Observations of other cores, such as those
studied here, will help determine to what extent the chemistry of
TMC-1 is representative of typical dense cores.

Figure~\ref{fig:spectra} shows the radio K-band spectra of the Chamaeleon
MMS1 and C2 cores taken with the Canberra DSN telescope. The main
individual lines detected, including those of HC$_3$N, HC$_5$N, c-C$_3$H$_2$,
C$_4$H, CCS, C$_3$S, NH$_3$, and the rare isotopologues $^{15}$NH$_3$
and c-C$_3$HD, are shown in more detail in Figures~\ref{fig:spemms1}
and \ref{fig:spec2}. The line parameters derived from Gaussian and
hyperfine-structure (HFS) fits are listed in Tables~\ref{tab:mms1} and
\ref{tab:c2}. Below we derive the molecular column densities and
relative fractional abundances of the detected molecules, as well as
upper limits for selected molecular species.

\begin{table}
\begin{center}  
\caption{Physical properties of the Chamaeleon MMS1 and C2 cores.} 
\label{tab:param}
\begin{tabular}{ccc}
\hline \hline 
  \rule[-3mm]{0mm}{8mm}  & MMS1 & C2 \\
\hline 
  $T_{d}$ (K) & 13.2 & 13.0 \\
  $T_{g}$ (K) & 8.5--10.9 & 8.5--10.9 \\
  $N_{\rm H_2}$ (cm$^{-2}$) & $4 - 17 \times 10^{22}$ & $2.5 - 10 \times 10^{22}$ \\
  $n_{\rm H_2}$ (cm$^{-2}$) & $3 - 14 \times 10^{5}$ & $2 - 8 \times 10^{5}$ \\
\hline
\end{tabular}
\end{center}
Note: Physical parameters are mean values within the 45\arcs\ DSS-43 beam. 
\end{table}

\subsection{Densities and temperatures of the Chamaeleon
  cores}\label{sec:dentemp} 

Calculations of molecular column densities require prior knowledge of
the gas temperature (local thermodynamical equilibrium, LTE,
calculations), or temperature and density (radiative transfer models,
such as the large velocity gradient, LVG, approach). These physical
parameters can be estimated from the existing dust continuum as well
as molecular data, given that multiple transitions covering a range of
excitation conditions are detected.

The two bottom panels in Figure~\ref{fig:spire} show dust spectral
energy distributions (SEDs) in the DSS-43 beam toward the MMS1 and C2
cores. Modified blackbody fits to the SPIRE and PACS 160~\mic\
\cmark surface brightness, \umark
$I_\nu = B_\nu (T_d) \times [1 - \exp(-\tau_{350} \times (350 \mu {\rm
  m} / \lambda)^\beta)]$, (shown as black curves in the bottom panels
of Fig.~\ref{fig:spire}) give average dust temperatures of \about 13 K
and a grain emissivity exponent $\beta=1.9$ in both cores. We exclude
the PACS 70~\mic\ data from the fit, as the emission may be
contaminated by a warmer dust component, distributed across the outer
layers of the cloud.

The H$_2$ column densities in the DSS-43 beam can be computed using the
formula
$N({\rm H_2}) = 2 a \rho R_{gd} / 3 m_{\rm H} \times \tau_{350} /
Q_{350}$ \citep{lis90}, where $a = 0.1\, \mu$m is the grain radius,
$\rho = 3$~g\,cm$^{-3}$ is the mean density, $R_{gd} = 100$ is the gas
to dust ratio, $m_{\rm H}$ is the hydrogen mass, and $Q_{350}$ is the
350~$\mu$m grain emissivity coefficient. Extrapolating the 125 $\mu$m
grain emissivity of \cite{hilde83} ($7.5 \times 10^{-4}$) with a
$\nu^2$ frequency dependence gives $Q_{350} = 1 \times 10^{-4}$,
corresponding to the grain mass opacity coefficient
$\kappa_{350} = 3 Q / 4 / a /\rho / R_{gd} = 0.025$~cm$^2$\,g$^{-1}$.
The values of $Q_{350}$ and $\kappa_{350}$ are highly uncertain and
values 4 times higher have been suggested for the Orion Molecular
Cloud (see the discussion in \citealt{lis98} and
\citealt{goldsmith97}).

The resulting H$_2$ column densities toward MMS1 and C2 are thus in
the range $N_{\rm H_2} = 4 - 17 \times 10^{22}$ and
$2.5 - 10 \times 10^{22}$~cm$^{-2}$, respectively. Assuming a
line-of-site depth equal to $\about 55^{\prime\prime}$ (as implied by
the 50\% contour of the 350~\mic\ emission in Fig.~\ref{fig:spire})
and a distance of 150~pc, the mean H$_2$ densities in the DSS-43 beam
toward MMS1 and C2 are $n_{\rm H_2} = 3 - 14 \times 10^5$ and
$2 - 8 \times 10^{5}$~cm$^{-3}$, respectively.

A wide range of column and volume densities in Cha~MMS1 has been
reported in the literature. \cite{kontinen00} derived a low H$_2$
column density of $1.2-1.6 \times 10^{22}$~cm$^{-2}$ and an average
H$_2$ density of $\sim 7 \times 10^4$~cm$^{-3}$ based on C$^{18}$O
observations using the SEST telescope (55\arcsec\ beam). However, this
value depends on the assumed fractional abundance of C$^{18}$O, which
may be lower than the canonical Taurus value \citep{frerking79}, if
partial freezeout occurs in cold, high density gas.

\cite{tennekes06} report a slightly higher H$_2$ column density range of
$1.9-4.0 \times 10^{22}$~cm$^{-2}$ based on 1.3~mm dust continuum
observations. Their spherical Monte Carlo model that fits SEST
observations of HCN isotopologues in Cha-MMS1 has densities ranging from
$2.3 \times 10^4$~cm$^{-3}$ at the outer 60\arcsec\ radius and
$1.4 \times 10^6$~cm$^{-3}$ at the center. \cite{cordiner2012} assume
an H$_2$ density of $10^6$~cm$^{-3}$ in the analysis of their ATNF Mopra
observations at 32--50 GHz (96--77\arcsec\ beam), while
\cite{belloche11} derive a peak column densities of 9.2 and
$2.7 \times 10^{22}$~cm$^{-2}$ in a 21\arcsec\ beam toward Cha-MMS1
and C2, respectively, and average gas densities in a 50\arcsec\
diameter aperture of $9.8 \times 10^5$ and $3.9\times10^5$~cm$^{-3}$,
respectively, based on 870~$\mu$m dust continuum observations using
the APEX telescope.

Observations of HC$_3$N and HC$_5$N for which theoretically derived
collisional rate coefficients are available can be used to provide an
independent estimate of the gas temperature and density. One advantage
to this approach is that their emission most likely arises from the
same region as other organic molecules. In Appendix~\ref{app:lvg} we show that our
HC$_3$N and HC$_5$N data in conjunction with SEST data of
\cite{kontinen00} in a similar beam are well reproduced by LVG models
with a kinetic temperature of $\about 8.5$~K and a density
$\gtrsim 3 \times 10^5$~cm$^{-3}$. This kinetic temperature is close to
the range 7.1--7.2~K derived by \cite{cordiner2012} based on Mopra
observations of HC$_3$N and HC$_5$N in a larger beam.

The average kinetic temperature of the gas within the DSS-43 beam can
also be constrained by observations of the inversion lines of ammonia
(Appendix~\ref{app:ammonia}). The rotational temperature of 10.9~K for
Cha-MMS1 derived in Appendix~\ref{app:ammonia} agrees with previous
ammonia observations, 12.1~K \citep{tennekes06}. These values are
higher than the temperature implied by LVG models of HC$_3$N and HC$_5$N
(Appendix~\ref{app:lvg}).

\subsection{LTE molecular column densities}

There are two possible approaches to converting the observed line
intensities to molecular column densities, depending on the
availability of collisional rate coefficients. For molecules with
calculated rate coefficients we can use these along with LVG radiative
transfer models to derive physical properties, such as density and
temperature of the molecular hydrogen, and the column density of the
trace molecule given sufficient transitions. 

Since for some molecules studied here collisional rates are not
available, we use the Weeds software
package\footnote{https://www.iram.fr/IRAMFR/GILDAS/doc/html/weeds-html/weeds.html}
which can perform a simple modeling of the observed spectra, under the
assumption of LTE. We use molecular spectroscopy data from the Cologne
Database for Molecular Spectroscopy (CDMS) catalog \citep{muller01,
  muller05} in the calculations.

For ammonia, we used the excitation temperatures derived from the
hyperfine structure fits (Sect.~3.2.5 below) while for other molecules we
used a value of 8.5~K, as derived from the LVG analysis of the HC$_3$N
and HC$_5$N data in Cha-MMS1 (Appendix~\ref{app:lvg}). We
fixed the source sizes for MMS1 and C2 cores to the mean values
derived above. We then adjusted the molecular column density as the
only parameter to match the observed line intensities reported in
Tables~\ref{tab:mms1} and \ref{tab:c2}. The resulting column densities
are listed in Table~\ref{tab:cden} along with selected column density
ratios in Table~\ref{tab:abun}. For molecules with multiple lines
detected, the column densities reported in Table~3 correspond to
averages of values derived from individual lines.

In the following subsections we discuss observations of individual
molecules detected in the DSS-43 spectra of Chamaeleon MMS1 and C2 and
compare their column densities and abundance ratios with reference
values for the cyanopolyyne peak in TMC-1 from \cite{gratier16}. In
general, column densities of molecular species in the Chamaeleon cores
are an order of magnitude lower than in TMC-1, \cmark in spite of the
higher H$_2$ column densities (Table~\ref{tab:param}) compared to 
$10^{22}$\,cm$^{-2}$ in TMC-1 \citep{gratier16}. \umark The high
molecular column densities make TMC-1 a preferred target for searches
for new molecular species. Notable exceptions are c-C$_3$H$_2$ and
NH$_3$, as discussed below.

\begin{table} 
\begin{center}  
\caption{LTE molecular column densities in the Chamaeleon MMS1 and C2
  cores compared to TMC-1.} 
\label{tab:cden}
\begin{tabular}{ccccc}
\hline \hline 
  \rule[-3mm]{0mm}{8mm}  Molecule & $N$(MMS1) & $N$(C2)
  & $N$(TMC-1) \\
  & (cm$^{-2}$) & (cm$^{-2}$) & (cm$^{-2}$) \\
\hline 
  HC$_3$N	& $1.7 \times 10^{13}$  & $8.1 \times 10^{12}$  & $2.3 \times 10^{14}$ \\
  HC$_5$N	& $3.7 \times 10^{12}$  & $2.2 \times 10^{12}$  & $5.9 \times 10^{13}$\\
  HC$_7$N	& $6.8 \times 10^{11}$  & $3.4 \times 10^{11}$  & $4.6 \times 10^{13}$\\ 
c-C$_3$H$_2$	& $3.1 \times 10^{13}$  & $1.5 \times 10^{13}$  & $1.9 \times 10^{13}$\\
c-C$_3$HD	& $6.8 \times 10^{12}$  & $3.7 \times 10^{12}$  & $5.1 \times 10^{12}$ \\
  CCS	        & $1.4 \times 10^{13}$  & $1.5 \times 10^{13}$  & $1.0 \times 10^{14}$\\
  C$_3$S	& $2.5 \times 10^{12}$  & $2.8 \times 10^{12}$  & $1.4 \times 10^{13}$\\
  C$_4$H	& $1.2 \times 10^{13}$  & $1.1 \times 10^{13}$  & $1.3 \times 10^{14}$\\ 
  $^{15}$NH$_3$	& $5.3 \times 10^{12}$ & $<8.0 \times 10^{12}$ \\
  NH$_3$        & $3.7 \times 10^{15}$ & $2.6 \times 10^{15}$   & $5.0 \times 10^{14}$\\
  c-C$_6$H$_5$CN  & $<8.9 \times 10^{11}$& $<6.0 \times 10^{11}$& $4 \times 10^{11}$ \\
\hline
\end{tabular}
\end{center}

Note: Assumed $T_{ex}=8.5$~K for all molecules except of ammonia
isotopologues for which $T_{ex}=7.6$~K and 5.5~K in Cha-MMS1 and Cha-C2,
respectively. TMC-1 values correspond to the cyanopolyyne peak
\citep{gratier16, mcguire2018}.
Ammonia column densities computed using Weeds have been multiplied by
1.038 (see Appendix~\ref{app:ammonia}) Upper limits are $3\sigma$.
\end{table}

\begin{table} 
\begin{center}  
\caption{Molecular column density ratios in the Chamaeleon MMS1 and C2
  cores compared to TMC-1.}
\label{tab:abun}
\begin{tabular}{cccc}
\hline \hline 
  \rule[-3mm]{0mm}{8mm}  Molecules & $R$(MMS1) & $R$(C2) & $R$(TMC-1) \\
\hline 
  HC$_3$N / HC$_5$N	& 4.7 & 3.8 & 4.0 \\
  HC$_7$N / HC$_5$N       & 0.18 & 0.16 & 0.78 \\
  C$_4$H  /HC$_5$N	& 3.2 & 5.0 & 2.1 \\ 
  CCS / HC$_5$N	        & 4.7 & 6.9 & 1.7 \\ 
  CCS / C$_3$S	        & 5.5 & 5.4 & 7.4 \\
 c-C$_6$H$_5$CN / HC$_5$N & $<0.24$ & $<0.28$ & 0.09 \\ 
\hline 
  c-C$_3$H$_2$ / HC$_5$N	& 8.3 & 7.1 & 0.32\\ 
  NH$_3$ / HC$_5$N	& 1000 & 1200 & 8.5 \\
  \cmark NH$_3$ / HC$_3$N	& 210 & 320 & 2.1$^a$ \umark \\
  NH$_3$ / c-C$_3$H$_2$   & 120 & 170 & 27 \\ 
\hline 
  c-C$_3$HD / c-C$_3$H$_2$	& 0.22 & 0.24 & 0.28 \\ 
  NH$_3$ / $^{15}$NH$_3$  & 69 & $>312$ \\ 
\hline
\end{tabular}
\end{center}

\cmark TMC-1 abundance ratios are from \citealt{gratier16}.
$^a$\cite{pratap97} derive independently a factor of 2 higher
NH$_3$/HC$_3$N abundance ratio of 5.7 at the cyanopolyyne peak in
TMC-1. When compared with \citealt{gratier16}, their value provides an
estimate of the uncertainty of the relative abundance determination.
\umark
\end{table}

\subsubsection{Cyanopolyynes}

A single rotational transition of HC$_3$N, $J = 2 - 1$, is within the
DSS-43 frequency range. The hyperfine splitting is clearly detected in
both cores, indicating optically thin emission. LVG models of the
DSS-43 and SEST observations are consistent with a kinetic temperature
of \about 8.5~K and a density  $\gtrsim 3 \times 10^{5}$ cm$^{-3}$. Three
rotational transitions of HC$_5$N, from $J=7-6$ to $9-8$, are detected
in both cores. HC$_7$N is also detected via stacking analysis (see Sect.~4).

Column densities of cyanopolyynes in the Chameleon cores are more than
an order of magnitude lower than those in TMC-1 \citep{gratier16}.
However, the relative abundance ratios HC$_3$N:HC$_5$N:HC$_7$N are
similar in both cores cores, $\sim 5$ (Table~\ref{tab:abun}). As a
reference, the HC$_3$N:HC$_5$N abundance ratio in TMC-1 is $\sim 4.0$,
comparable to the Chamaeleon values, but the HC$_5$N:HC$_7$N ratio is
lower, $\sim 1.3$ \citep{gratier16}. This difference may indicate that the
production of long carbon chains is less efficient in Chamaeleon
compared to TMC-1.

\subsubsection{Polyynes}

Three lines of C$_4$H are detected in both cores. Our column density
computations used the latest CDMS value of the dipole moment, 2.1~D,
from a quantum chemical calculation by \cite{oyama20}. For
comparison, we use the \cite{gratier16} C$_4$H column density in
TMC-1 scaled to a dipole moment of 2.1~D in Tables~\ref{tab:cden} and
\ref{tab:abun}. The resulting C$_4$H:HC$_5$N ratios in the Chamaeleon
cores are comparable to that in TMC-1.

\subsubsection{Sulfur species}

Lines of two sulfur-bearing species, CCS and C$_3$S are detected in
both cores. Compared to TMC-1, the CCS abundance appears slightly
enhanced compared to cyanopolyynes, while the CCS/C$_3$S ratio is
similar to that measured in TMC-1.

\subsubsection{Cyclopropenylidene isotopologues}

Two lines of c-C$_3$H$_2$ are detected within the DSS-43 frequency
range. The $2_{20} - 2_{11}$ line at 21.6 GHz is seen in absorption
against the CMB, as previously observed in TMC-1 \citep{madden89} and
consistent with LVG models. The
$1_{10} - 1_{01}$ line at 18.3 GHz is seen in emission and was used to
derive column density estimates. In contrast to cyanopolyynes, column
densities of c-C$_3$H$_2$ in the two Chamaeleon cores are comparable
to TMC-1. The resulting c-C$_3$H$_2$/HC$_5$N ratio is thus a factor of
\about 25 higher compared to TMC-1 (Table~\ref{tab:abun}).

A line of the deuterated isotopologue c-C$_3$HD is also detected with
a high S/N ratio. The resulting D/H ratio is the same in the two
cores, \about 0.23, and consistent with the TMC-1 value.
\cite{majumdar17} carried out a detailed study of c-C$_3$HD in the
solar type protostar IRAS16293-2422 and also derived a high deuterium
fraction of 0.14, an order of magnitude higher than the value
predicted by their chemical model. A possible explanation is that
current astrochemical models appear to over-predict the abundance of
c-C$_3$H$_2$ \citep{agundez13, sipila16}.

\subsubsection{Ammonia isotopologues}

The hyperfine structure (HFS) is clearly detected in the (1,1) and
(2,2) lines in both sources. HFS fits to the ammonia inversion line
spectra, carried out using the IRAM CLASS software
(Figure~\ref{fig:ammonia}), indicate that the NH$_3$ (1,1) line is
optically thick in both Cha-MMS1 and C2 (optical depths of the main
hyperfine component equal to 7.6 and 2.9, respectively,
Tables~\ref{tab:mms1} and \ref{tab:c2}). The (2,2) line is optically
thin, within the fit uncertainties, and the (3,3) line is only
detected in Cha-MMS1. The excitation temperatures within the $K=1$
rotational ladder can be derived directly from the HFS fits
\footnote{See Sect.~2.5 of the GILDAS CLASS user manual,
  https://www.iram.fr/IRAMFR/GILDAS/doc/pdf/class.pdf.}. From the
optically thick spectra of the (1,1) inversion transitions, corrected
for the source coupling, we derive excitation temperatures of 7.6 and
5.5~K, for Cha-MMS1 and Cha-C2, respectively.

In addition to the excitation temperature, the two K-ladders of
ammonia can be used to derive the kinetic temperature. In
Appendix~\ref{app:ammonia} we discuss the derivation of the ammonia
rotational temperature, taken as a measure of the kinetic temperature
of the gas and the corresponding correction to the molecular column
density derived from LTE analysis. We derive kinetic temperatures of
10.9 and 11.1~K in Cha-MMS1 and Cha-C2, respectively.

The ammonia column densities in the two Chamaeleon cores are an order
of magnitude higher than in TMC-1. We note that our Cha-MMS1 value of
$3.6 \times 10^{15}$~cm$^{-2}$ is in good agreement with the value of
$1.4 \times 10^{15}$~cm$^{-2}$ derived by \cite{tennekes06}, after
correcting for the difference in beam filling factors between DSS-43
and a much larger 80$^{\prime\prime}$ beam of the Parkes telescope.
The resulting NH$_3$/HC$_5$N abundance ratio is a factor of \about 125
higher in the Chamaeleon cores compared to TMC-1. The NH$_3$ abundance
enhancement with respect to cyanopolyynes in Chamaeleon compared to
TMC-1 is thus a factor of \about 5 higher compared to that in
c-C$_3$H$_2$.

Because of the very high ammonia column density, the (1,1) line of
$^{15}$NH$_3$ is also detected toward Cha-MMS1 at a S/N ratio of 8.
LTE models using Weeds imply a high $^{14}$N/$^{15}$N isotopic ratio
of \about 690 in ammonia. We note that we assumed the same source size
of 52$^{\prime\prime}$ in the computations of the $^{14}$NH$_3$ and
$^{15}$NH$_3$ column densities. Given the high optical depth of the
$^{14}$NH$_3$ (1,1) line, it is possible that the effective source
size is larger compared to $^{15}$NH$_3$. A 50\% increase in the
$^{14}$NH$_3$ source size would reduce the column density by a factor
of 1.3, bringing the $^{14}$N/$^{15}$N isotopic ratio in ammonia down
to of \about 525.

As discussed by \cite{lis10}, nitrogen displays the largest isotopic
variations in the solar system after hydrogen, typically explained by
mixing of various proto- solar or pre-solar reservoirs. Earth, Mars,
Venus, and most primitive meteorites have nitrogen isotopic ratios
within 5\% of the terrestrial atmospheric value, $^{14}$N/$^{15}$N =
272 (see \cite{marty09}). However, the proto-solar nebula was poorer
in $^{15}$N, with $^{14}$N/$^{15}$N = 450, as evidenced by infrared
and in situ measurements in the Jupiter atmosphere
($530^{+380}_{-170}$ , ISO, \citealt{fouchet00}; $435 \pm 57$,
Galileo, \citealt{owen01}; $448 \pm 62$, from \emph{Cassini}, \citealt{abbas04};
$450 \pm 106$, \emph{Cassini}, \citealt{fouchet04}) and solar wind
measurements ($442 \pm 131$, 2$\sigma$, from \emph{Genesis}, \citealt{marty09}).
The proto-solar $^{14}$N/$^{15}$N ratio is in agreement with the local
interstellar medium (ISM) value ($450 \pm 22$, \citealt{wilson94}; or
$414 \pm 32$ at the birth place of the Sun, \citealt{wielen97}).

As discused by \cite{redaelli23} and references therein, nitriles are
often enriched in $^{15}$N with typical $^{14}$N/$^{15}$N abundance
ratios of 140--460, while N$_2$H$^+$ appears instead deficient in
$^{15}$N with isotopic ratios 580--1000. Using GBT, \cite{lis10}
measured $^{14}$N/$^{15}$N ratios of $334 \pm 50$ and $344 \pm 173$
(3$\sigma$) in ammonia in Barnard 1 and NGC~1333, respectively. In a
recent study \cite{redaelli23} derived a low isotopic ratio of
$210 \pm 50$ toward NGC~1333 IRAS4A and $390 \pm 40$ toward the center
of L1544.

\cite{redaelli23} argue that ammonia ices are enriched in
$^{15}$N, leading to a decrease in the $^{14}$N/$^{15}$N ratio when
the ices are sublimated back into the gas phase for instance due to
the temperature rise in protostellar envelopes. The high
$^{14}$N/$^{15}$N in Cha-MMS1 may suggest that ice sublimation is
not a dominant process within the DSS-43 beam.

\begin{figure*}
\centering
\includegraphics[trim=2.5cm 3.0cm 4.5cm 1.5cm, clip=true, width=0.8\textwidth,angle=0]{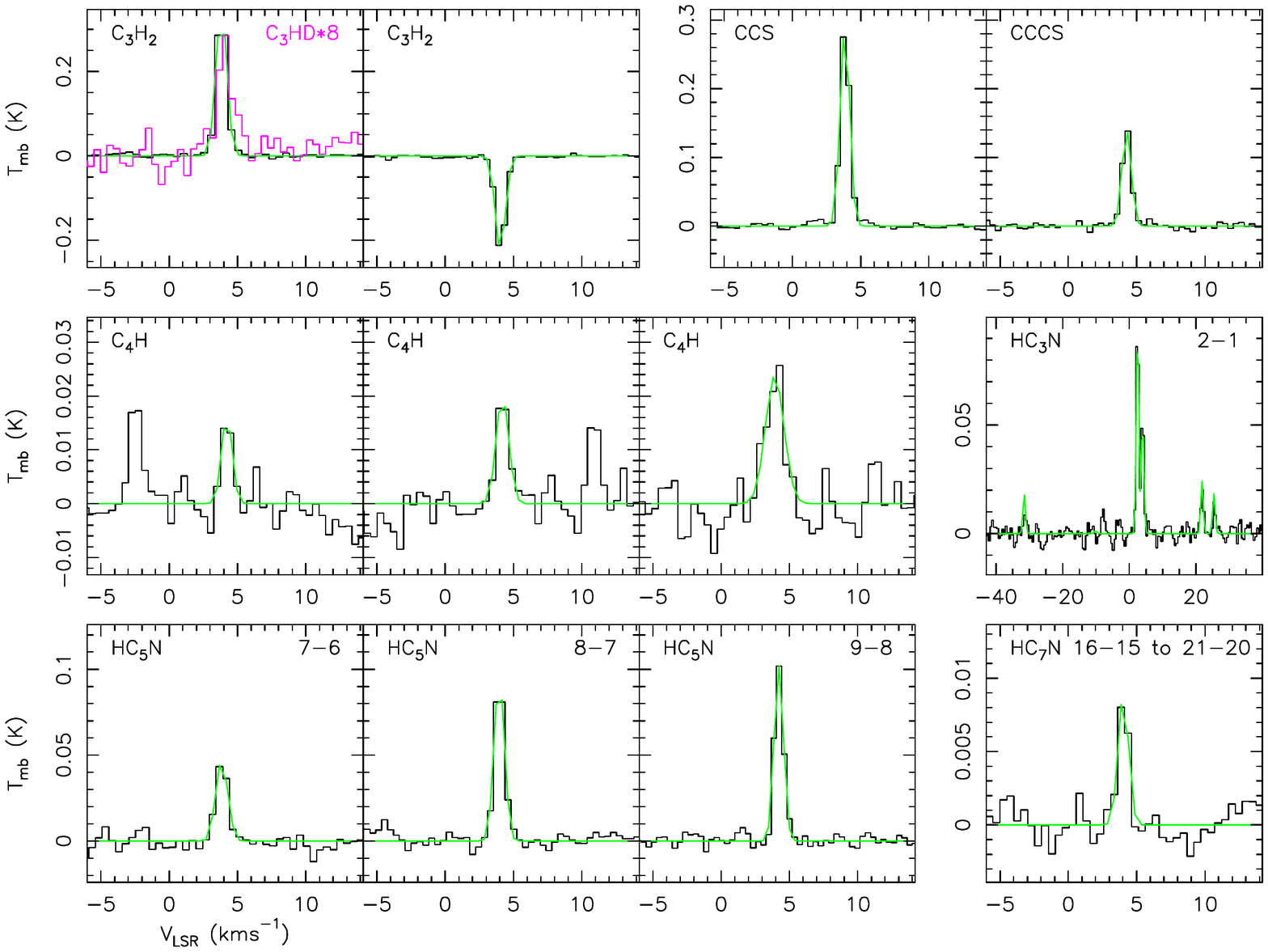} 
\caption{Spectra of molecular lines other than ammonia detected in
  Chamaeleon C2 (black and magenta histograms) with fits 
  shown in green.}
\label{fig:spec2}
\end{figure*}

\begin{figure*}
\sidecaption
\includegraphics[trim=2.0cm 6.5cm 1.5cm 2.5cm, clip=true, width=12cm]{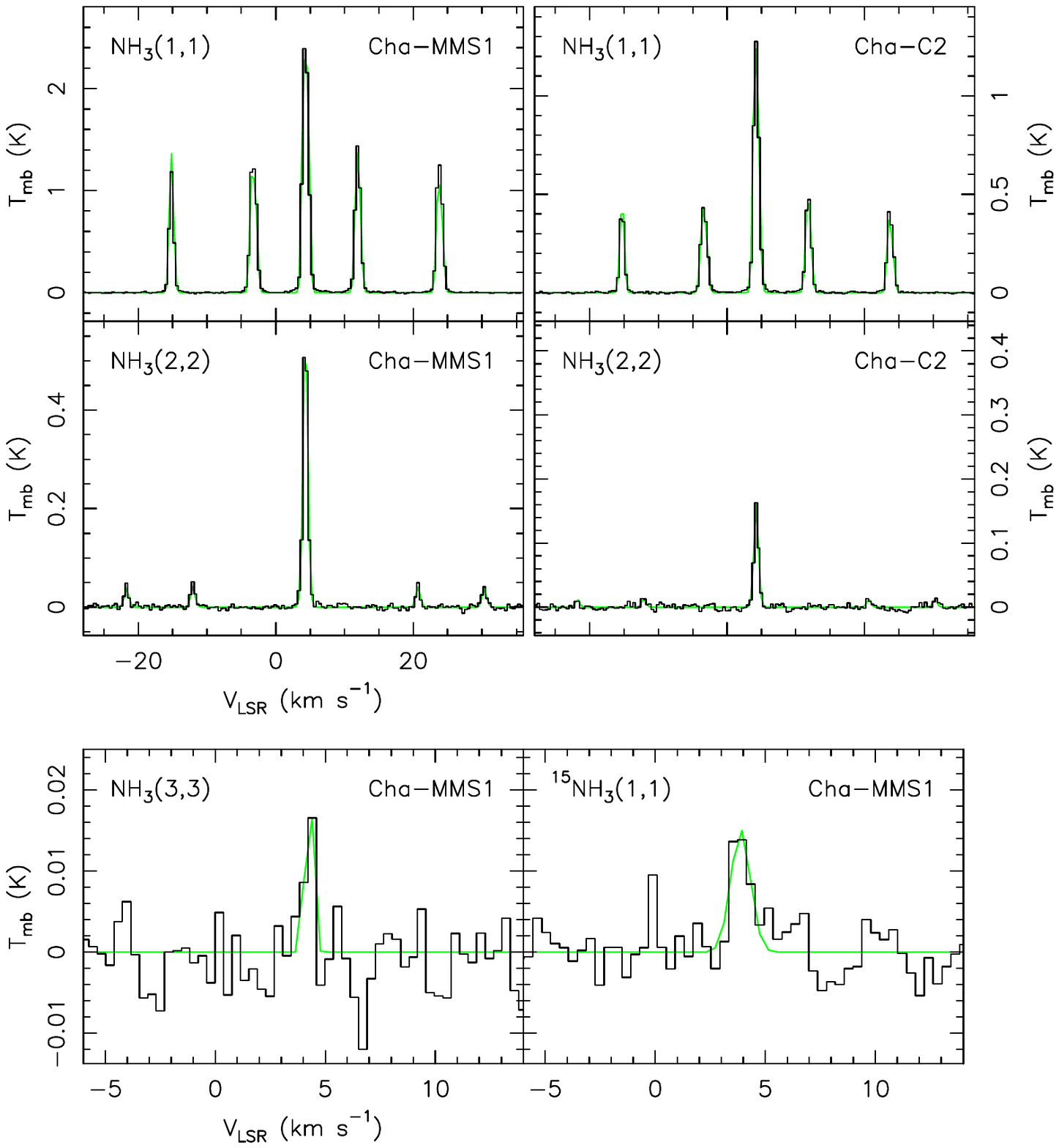} 
\caption{Spectra of the detected ammonia inversion lines in
    Chamaeleon MMS1 and C2. The transition and source are labeled in
    each panel. The observational data are shown as black histograms,
    and spectral fits are shown in green.}
\label{fig:ammonia}
\end{figure*}

\begin{figure*}
  \centering
\includegraphics[trim=2cm 2cm 6cm 2cm, clip=true,
width=0.8\textwidth,angle=0]{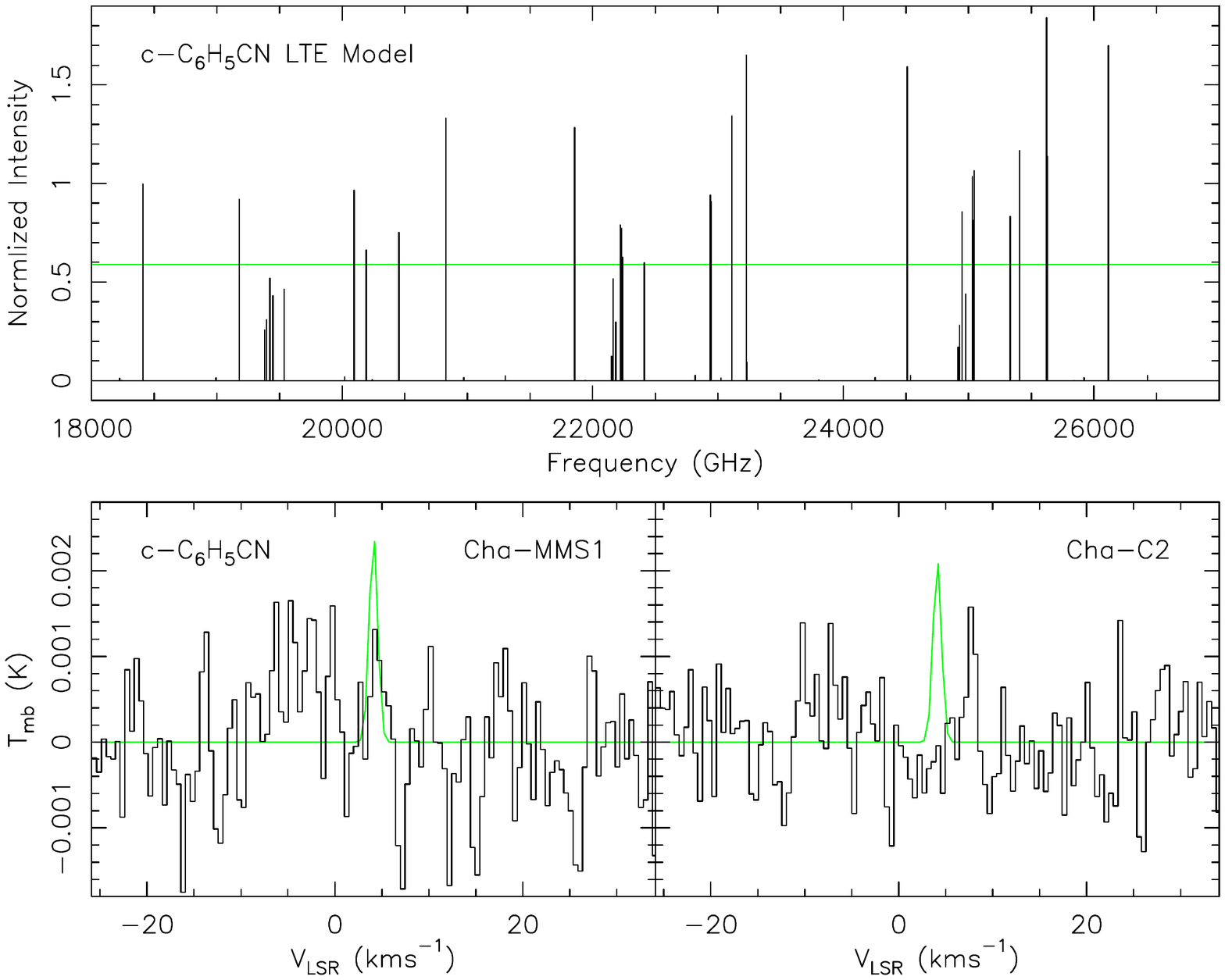}  
\caption{(Upper) LTE model spectrum of benzonitrile in the DSS-43
  frequency range for a temperature of 8.5 K. Line intensities are
  normalized to the 18.41~GHz transitions. Transitions above the green
  horizontal lines are included in the stacking analysis. (Lower)
  Stacked spectra of benzonitrile toward Cha-MMS1 and Cha-C2
  normalized to the 18.41~GHz line. Green curves are $3\sigma$ upper
  limits LTE models corresponding to column densities reported in
  Table~\ref{tab:cden}.}
\label{fig:benzo}
\end{figure*}

\section{Detections and limits obtained using line stacking analysis}
\label{sec:stacking}

Spectral line stacking of different transitions arising from the same
species has been proposed to overcome the low signal-to-noise ratio of
these faint signals across wavelength regimes \citep{chen13,
  lindroos16}, and has already shown success in detecting new species
at radio frequencies \citep{loomis18, walsh16}. By aligning signals
from multiple weaker transitions and considering their collective
effect, the signal-to-noise ratio can surpass a detection threshold
that individual lines might fall below. \cite{loomis21} introduced a
detection method specialized for sparse data, combining Markov Chain
Monte Carlo (MCMC) inference with spectral stacking. Their version of
the MCMC code available on GitHub is specific to GBT observations of
TMC-1. We generalized this code for use with other sources as
described in Appendix~\ref{app:stacking} and applied this revised MCMC
code to our DSS-43 spectra of HC$_7$N in the Chamaeleon molecular cloud.

\subsection{Longer cyanopolyynes}

We detected HC$_7$N with a high S/N ratio by stacking its rotational
transitions from $J = 16-15$ to $21-20$. Individual spectra are
normalized to the brightest line, as predicted by the MCMC model, and
weighted by 1/$\sigma^2$ in the final average. This approach leads to
a robust $9.1 \sigma$ detection of HC$_7$N in Cha-MMS1 and a
$7.5 \sigma$ detection in Cha-C2. The resulting spectra are shown in
Figs.~\ref{fig:spemms1} and \ref{fig:spec2}, lower-right panels and
the column densities and abundance ratios are reported in
Tables~\ref{tab:cden} and \ref{tab:abun}, respectively.

\subsection{Upper limit for benzonitrile}

The DSS-43 frequency range covers 350 rotational transitions of
benzonitrile covering a wide range of line intensities and upper level
energies. Figure~\ref{fig:benzo} (upper panel) shows a template LTE
model of the benzonitrile emission for a temperature of 8~K,
normalized to the intensity of the 18.41~GHz line. Transitions above
the green horizontal line, which corresponds to 32\% of the intensity
of the brightest line, are included in the stacking analysis. This
cutoff is arbitrary, but it corresponds to a factor of 10 increase in
the integration time to reach the same S/N ratio compared to the
brightest line. Including weaker line lines in the analysis does not
improve the S/N ratio in the stacked spectrum.

We extracted spectra of all lines above the cutoff defined above and
scaled them to the intensity of the 18.41~GHz line using the LTE
template. We then averaged all spectra using $1/\sigma^2$ weighting.
The resulting spectra toward Cha-MMS1 and Cha-C2 are shown in
Figure~\ref{fig:benzo} (lower panel). No benzonitrile emission is
detected in either source. The corresponding $3\sigma$ upper limits
for the column density are listed in Table~\ref{tab:cden}, and the
resulting model spectra are shown as green curves in
Figure~\ref{fig:benzo}. The $3 \sigma$ upper limits for the
benzonitrile column density achieved in the Chamaeleon sources are a
factor of 2 higher than the value derived for TMC-1
\citep{mcguire2018} and the corresponding upper limits for the
relative abundance of benzonitrile with respect to HC$_5$N are a factor
of 3 higher than the TMC-1 value.

\subsection{Unidentified features}

A single $\about 7.5 \sigma$ unidentified feature is seen in the Cha-C2
spectrum at 22.902~GHz. The feature is not present in the MMS1
spectrum.

\section{Discussion}
\label{sec:discussion}

One general result from our study is that column densities of most
molecules detected in the Chamaeleon sources are comparatively lower
than those in TMC-1. Interestingly, the average temperatures of both
Chamaeleon sources are comparable to that of TMC-1, approximately 10~K
(an average of the gas and dust temperatures), while the average
density of our sources is 10 times or more higher than that of TMC-1.

\cmark While the cynaopolyyne peak in TMC-1 is a particularly well
studied starless core, the presence of molecular abundance gradients
across TMC-1 is well established in the literature. \cite{pratap97}
compare relative abundances of several molecules toward the
cynaopolyyne, ammonia, and SO peaks in TMC-1. Of particular interest
for the results presented here is the NH$_3$/HC$_3$N ratio,
determined to be 5.8, 16.7, and 25.4 at the three positions,
respectively. \cite{hirahara92} attribute such variations to
differences in the chemical evolutionary stage, with carbon-chain
molecules produced in early stages and ammonia in late stages. While
the abundance ratios at the ammonia and SO peaks in TMC-1 are higher
than that at the cynaopolyyne peak, they are still significantly lower
than the values derived here for the Chamaeleon cores
(Table~\ref{tab:abun}). In the \cite{hirahara92} scenario, this would
imply that the Chamaeleon cores are characterized by late stage
chemistry.

\cite{law18} studied carbon chain molecules toward 16 embedded
low-mass protostars. The median molecular column densities of HC$_3$N,
HC$_5$N, CCS, and C$_3$S in their sample are a factor of 5--12 lower
that those in the Cha-MMS1 core, while the C$_4$H median column
density is a factor of 1.4 higher. The resulting median
HC$_3$N/HC$_5$N and CCS/HC$_5$N abundance ratios, $\sim 10$, are about
a factor 2 higher than the Chamaeleon values, while the CCS/C$_3$S
median abundance ratio, $\sim 25$, is a factor of 5 higher. The median
C$_4$H/HC$_5$N ratio, $\sim 110$, is a factor of 30 higher than the
Chamaelon values (see Table 5 of \citealt{law18}). \umark

Several additional factors may explain the differences in the observed
column densities between these sources. First, in higher-density
environments such as those in the Chameleon sources, molecules likely
freeze out onto dust grains more efficiently, as the timescale for the
freeze-out is inversely proportional to the gas density \citep{seo19}.
Second, lower-density regions like TMC-1 may have longer chemical
timescales, allowing molecular species to persist for extended periods
and resulting in higher gas-phase column densities \citep{majumdar15}.
Third, lower-density cores may allow greater penetration of ambient UV
photons, facilitating photodesorption of molecules from grain surfaces
and leading to higher abundances \citep{oberg07}. Fourth, variations
in cosmic-ray ionization rates between these sources could
significantly influence gas-phase chemistry, contributing to the
differences in the observed column densities \citep{taniguchi24}.
\citep{seo19}. \cmark It is also possible that these cores have
different ages, as well as different initial elemental abundances,
leading to different C/O ratios compared to TMC-1, is also a major
possibility \citep{taniguchi24}. For example, \cite{loison14} reported
that several carbon-chain groups (such as C$_n$, C$_n$H, C$_n$H$_2$,
C$_{2n+1O}$, C$_n$N, HC$_{2n+1}$N, C$_{2n}$H$^-$, and C$_3$N$^-$) show
a strong dependence on the assumed C/O ratios and evolutionary stage.
In particular, for carbon chains gas-phase chemistry dominates in the
early stages (~10$^5$ years), whereas depletion becomes significant in
the later stages. \umark

Determining which of these processes dominates in our case will
require future studies employing advanced chemical models, including
isotope chemistry observed in our samples. This will be a focus of a
future investigation.

\section{Summary}
\label{sec:conclusion}

Here we have extended the survey for organics in the southern hemisphere
to 1.3 cm by observing two cores in the Chamaeleon complex using
NASA's Deep Space Network antenna in Canberra, Australia, over the
frequency range of 18 to 25 GHz. We surveyed
the class 0 protostar Cha-MMS1 and the prestellar core Cha-C2, which
represent two stages in the evolution of dense cores. We used
the detections of ammonia, cyanopolyynes, and far-infrared dust
continuum to characterize the density and temperature in the
Chamaeleon cores and calculate the molecular column densities and
their relative ratios.

The main results can be summarized as follows:
\begin{itemize}
\item Several molecules are detected in both cores including
  HC$_3$N, HC$_5$N, C$_4$H, CCS, C$_3$S, NH$_3$, and c-C$_3$H$_2$. A
  longer cyanopolyyne, HC$_7$N, is detected with high confidence via
  spectral stacking analysis.

\item While molecular column densities in the
  two Chamaeleon cores are typically an order of magnitude lower
  compared to the cynaopolyyne peak in TMC-1, the molecular abundance
  ratios are in general agreement with the  TMC-1 values. The two
  exceptions are c-C$_3$H$_2$, which is enhanced by a factor of \about
  25 with respect to cyanopolyynes in the Chamaeleon cores, and
  ammonia, which is enhanced by a factor of \about 125.

\item A deuterated isotopologue c-C$_3$HD is detected in both cores,
  with a high D/H ratio of $\about 0.23$ in c-C$_3$H$_2$, in general
  agreement with observations of other sources, such as TMC-1 or
  IRAS1629-2422. 

\item A rare isotopologue of
  ammonia, $^{15}$NH$_3$, is also detected in Cha-MMS1 suggesting a
  high $^{14}$N/$^{15}$N ratio of $\about 690$ in ammonia. However,
  this ratio may be artificially enhanced due to the high optical
  depth of the $^{14}$NH$_3$ (1,1) line, which increases the effective
  source size.

\item The ring molecule benzonitrile, a tracer for the non-polar
  molecule benzene, is not detected in either Chamaeleon core. The
  $3 \sigma$ upper limits for the benzonitrile column density achieved
  are about a factor of 2 higher than the TMC-1 value
  \citep{mcguire2018} and the resulting upper limits for the relative
  abundance of benzonitrile with respect to HC$_5$N are a factor of 3
  higher than that measured in TMC-1.
  
\end{itemize}

Our results suggest that the chemical composition of the two
Chameleon cores is different in several aspects from that of TMC-1.
However, the chemical composition of the observed
organics in the two Chameleon cores are similar to each other despite
representing different stages of core evolution.  This result suggests
there may be a steady state solution during this period and it will
require comparison to earlier and later stages of cores to reveal the
relationship between chemical composition and age.
Ongoing DSS-43 observations of additional sources will determine
whether these conclusions are generally applicable to other southern star
forming cores. 

\begin{acknowledgements}
  This research was carried out at the Jet Propulsion Laboratory,
  California Institute of Technology, under a contract with the
  National Aeronautics and Space Administration (80NM0018D0004) and
  funded through the internal Research and Technology Development
  program. We thank Steve Lichten, Joe Lazio, and the DSN staff for
  their support and assistance with the DSS-43 observations, \cmark and
  an anonymous referee for helpful comments. \umark L.M. acknowledges
  financial support provided by DAE and the DST-SERB research grant
  (MTR/2021/000864) from the Government of India.
\end{acknowledgements}

\begin{appendix}

\section{LVG analysis of HC$_3$N and HC$_5$N}\label{app:lvg}

We use a large velocity gradient (LVG) model of HC$_3$N and HC$_5$N
spectra to determine the best fit densities and temperatures for the
cores where organic molecules are detected. For both species
theoretical calculations exist for the collisional rate coefficients
and there are sufficient number of spectral lines detected from
different transitions to constrain the solutions. We use the offline
version of the RADEX code \citep{vandertak07} to calculate intensities
$I$(K \kms), opacity $\tau$, and level population as a function of
density, kinetic temperature, and column density. We consider a range
of $T_k$, $n$(\hh), and column density N(HC$_n$N) suggested by earlier
observations of the Chamaeleon cores to find a good fit to the data.
We use the line widths listed in Table~\ref{tab:mms1} and a background
temperature $T_{bg}$ =2.73K. For HC$_3$N we use collisional rate
coefficients from \cite{faure16} which are available from the Leiden
Atomic and Molecular
database\footnote{https://home.strw.leidenuniv.nl/~moldata/} while
those for HC$_5$N are from \cite{lique25}. Given the low gas temperatures,
in both cases we assume that
hydrogen is para-\hh. The resulting intensities were compared to the
DSN and SEST data for HC$_3$N. The DSN and SEST beam widths are
similar, $\sim$45$\arcs$ and 55$\arcs$, respectively and comparable to
the size of the cores, as discussed in Section~\ref{sec:dsn}. We
exclude the 7~mm Mopra observations of HC$_3$N and HC$_5$N
\citep{cordiner2012} because its beam size, 96$\arcs$ to 77$\arcs$, is
much larger than those of the DSN and SEST, and larger than the source
size. These differences in angular resolution introduces uncertain
filling factor corrections for beam dilution and beam coupling.

\begin{figure}
\centering

\includegraphics[trim=2cm 3cm 5cm 2cm, clip=true,
width=\columnwidth,angle=0]{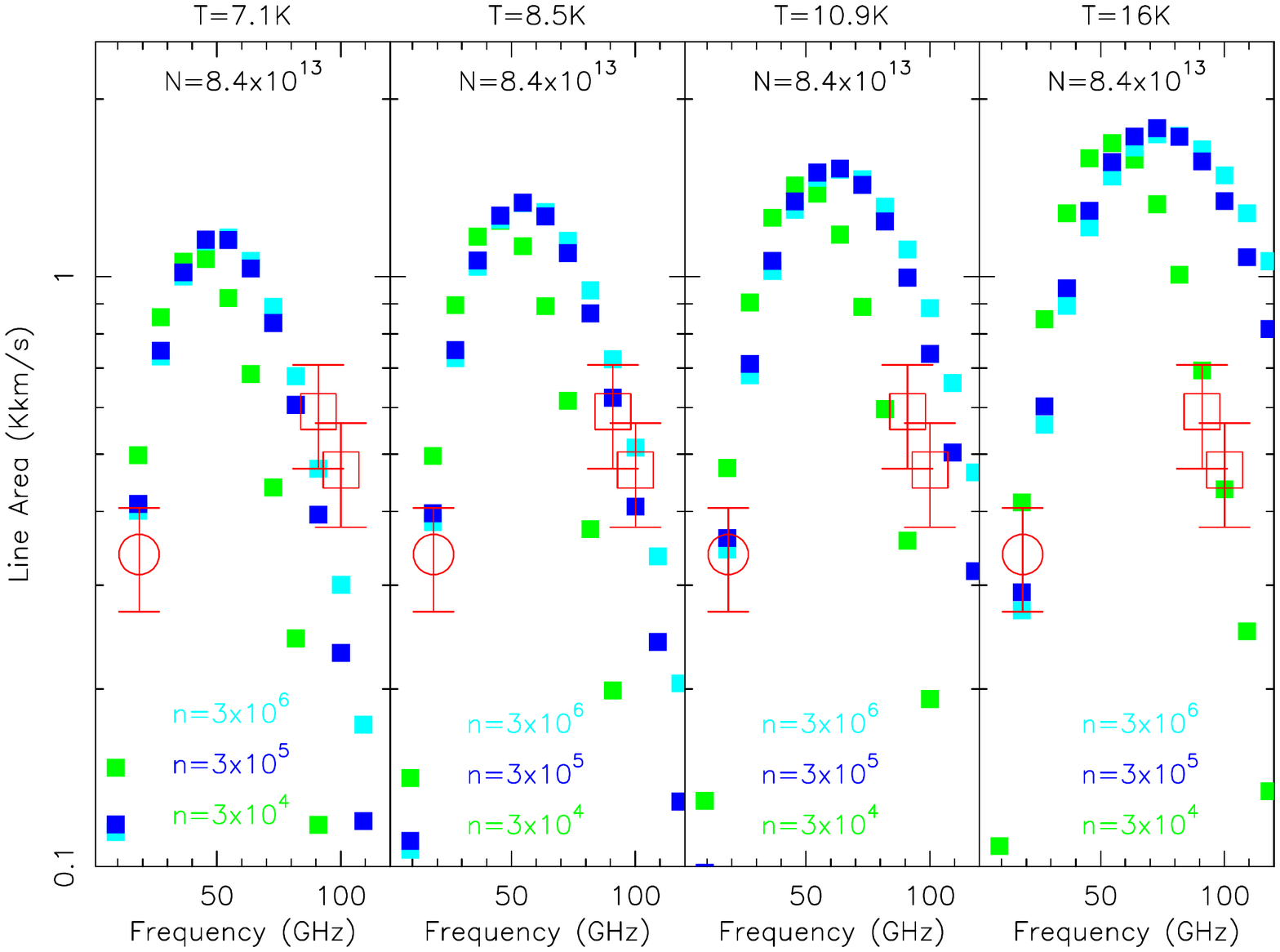}  

\caption{LVG models of the HC$_3$N emission in Cha-MMS1
  for different densities (filled cyan, blue, and green squares, as
  labeled). The four panels correspond to kinetic temperatures of 7.1,
  8.5, 10.9, and 16~K, left to right, respectively. The open red
  circle shows our DSN observation while the red squares are SEST
  observations of \cite{kontinen00}. The error bars correspond to
  a typical 20\% absolute calibration uncertainty.}
\label{fig:lvghc3n}
\end{figure}

Figure~\ref{fig:lvghc3n} shows results for HC$_3$N intensities for
densities, $n$(H$_2$) = 3\tim10$^4$, 3\tim10$^5$, and 3\tim10$^6$
\pccm\ which cover the range of densities derived from dust emission
and C$^{18}$O, as discussed in Sect.~~\ref{sec:dentemp}. We also consider
four values of kinetic temperatures, as follows: $T_k =7.1$~K is the
LTE value derived by \cite{cordiner2012} for HC$_3$N and HC$_5$N,
8.5~K is our best solution for HC$_3$N and HC$_5$N, 10.9~K is derived
from our ammonia data, and 16~K was chosen to study the impact of a
higher temperature on the distribution of intensities. The solution
for each $T_k$ is shown in a separate panel and the DSN observed
intensity is plotted as a circle and the SEST ones as squares. The
best fit solution is for $T_k = 8.5$~K and column density N(HC$_3$N)
=8.4\tim10$^{13}$ \pscm. There is little difference between the
solutions at densities $3 \times 10^{5}$ and $3 \times 10^6$ \pccm. At
densities above $3 \times 10^5$ cm$^{-3}$ the transitions considered
here are approaching thermalization. In fact, low-frequency HC$_5$N
lines are fully thermalized at densities as low as $10^5$~cm$^{-3}$,
as suggested by their critical densities. In the LVG solution all the
observed HC$_3$N lines have an opacity $\tau$ < 0.2 and the excitation
temperatures range from 7.4~K to 8.6~K consistent with the lines
approaching thermal equilibrium for the case $T_k = 8.5$~K. We note
that a solution with a low density of $3 \times 10^4$~cm$^{-3}$ and a
higher temperature of 16~K is also consistent with the observations.
However, these temperature and density values are inconsistent with
other the values suggested by other tracers 
(Sect.~\ref{sec:dentemp}).

\begin{figure}
\centering

\includegraphics[trim=2cm 3cm 5cm 2cm, clip=true,
width=\columnwidth,angle=0]{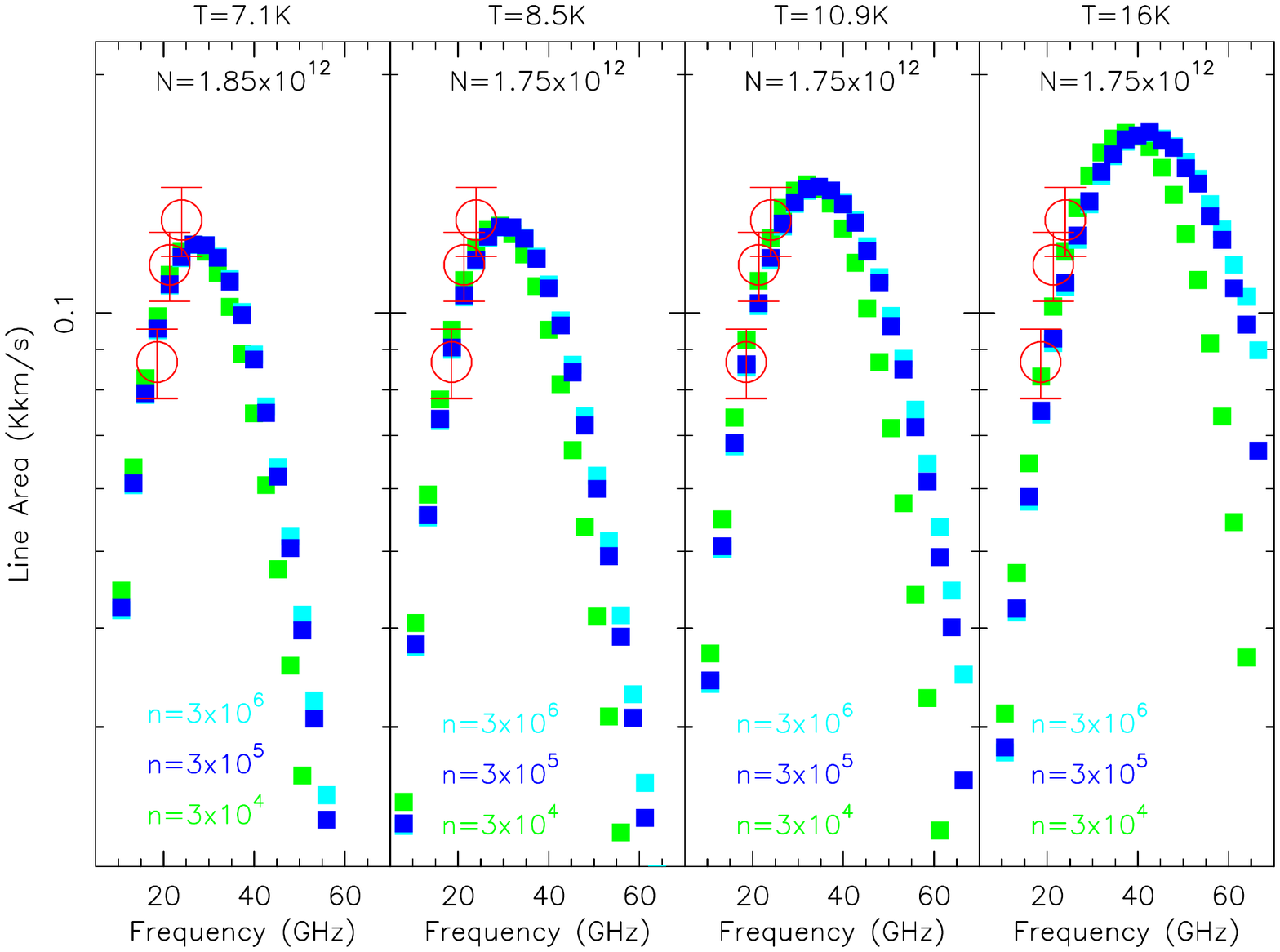}  

\caption{LVG models of the HC$_5$N emission in Cha-MMS1 for different
  densities (filled cyan, blue, and green squares, as labeled). The
  four panels correspond to kinetic temperatures of 7.1, 8.5, 10.9,
  and 16~K, left to right, respectively. The open red circles show
  our DSN observations. The error bars correspond to a typical 10\%
  relative calibration uncertainty.}
\label{fig:lvghc5n}
\end{figure}

For HC$_5$N we ran an LVG analysis over the same range of $T_k$ and
$n$(H$_2$) and the results are shown in Figure~\ref{fig:lvghc5n}. The column
density N(HC$_5$N) = $ 1.75 \times 10^{12}$ \pscm\ for the best fit
temperature $T_k = 8.5$~K, and is very similar for the other three
values of $T_k$. The intensities of the three lines detected with the
DSN are shown as red circles. The constraints on physical parameters
from HC$_5$N is less stringent than for HC$_3$N as SEST did not detect
any HC$_5$N lines. We have excluded the HC$_5$N data from Mopra
\citep{cordiner2012} for the same reasons discussed above for HC$_3$N.
Without enough contrast in excitation conditions we can only constrain
$T_k$ from 8.5~K to 16K, and $n$(H$_2$) to be greater than
$3 \times 10^4$ cm$^{-3}$. However, the column density is insensitive
over the likely range of temperatures and densities considered here,
as well as those indicated by the HC$_3$N analysis, and that is the
key parameter needed for interpreting chemical abundances.

\section{Ammonia rotational temperatures and column
  densities}\label{app:ammonia}

In the case of ammonia, the column density calculation is complicated
by the presence of two different temperatures describing the
rotational level population: the excitation temperature $T_{ex}$
within a given $K-$ladder, and the rotational temperature $T_{rot}$
describing the relative populations of the metastable rotational
levels at the bottom of each $K-$ladder. In the absence of allowed
radiative transitions between different $K-$ladders, the latter can be
taken as a measure of the gas kinetic temperature.

In spite of this complication, we can use the excitation temperature
derived from the HFS fit to determine the ammonia column densities in
the lowest metastable level of each $K-$ladder. These values can then
be used to determine the rotational (kinetic) temperature of the gas.

The upper level molecular column density of a rotational
transitions within the telescope beam can be computed from the
observed integrated line intensity of the transitions using the
standard formula (see, e.g., \citealt{lis02})
\begin{equation}
  N_{u} = {{8 \pi k \nu^2} \over {h c^3 A_{ul}}} ~ 
  {1 \over {1 - [J_\nu(T_{bg})/J_{\nu}(T_{ex})]}} ~ \int T_{R} dv~,
\end{equation}

\noindent where $\nu$ is the line frequency, $T_{ex}$ is the
excitation temperature, $A_{ul}$ is the Einstein
spontaneous emission coefficient, $E_u$ is the upper level energy,
$J_\nu(T)= h\nu/k/(e^{h\nu/kT} -1)$ is the radiation temperature of a
blackbody at a temperature $T$, $T_{bg}$ is the cosmic background
temperature (2.7~K), and $\int T_{R} dv$ is the opacity and beam
efficiency corrected integrated line intensity.

The ratio of the populations of the upper and lower levels of the inversion
transitions at the bottom of each $K-$ladder is given by
$N_l/N_u=g_l/g_u\exp((E_u-E_l)/kT_{ex}$. Therefore, the population in the lowest
metastable level of each $K-$ladder can be computed as
\begin{equation}
  N_{l} = {g_l \over g_u} e^{(E_u-E_l)/kT_{ex}}
  {{8 \pi k \nu^2} \over {h c^3 A_{ul}}} 
  {1 \over {1 - [J_\nu(T_{bg})/J_{\nu}(T_{ex})]}} \int T_R dv~.
\end{equation}

\noindent Here, the excitation temperature is that derived from the
HFS fit and the resulting column densities for Cha-MMS1 and Cha-C2 are
listed in Table~\ref{tab:ammomms1}.

The rotational temperature can then be computed from the populations
of two metastable levels at the bottom of different $K-$ladders,
denoted $i$ and $j$, using the formula
\begin{equation}
  N_i/N_j = {g_i \over
    g_j} \exp(-{{E_i-E_j} \over kT_{rot}})~.
\end{equation}

\noindent From the lower-level column densities of the (1,1) and (2,2)
transitions reported in Table~\ref{tab:ammomms1} we derive rotational
temperatures of 10.9~K both in Cha-MMS1 and Cha-C2.

The total NH$_3$ column densities computed by the Weeds package use
the partition functions computed assuming that all levels are populated at the
assumed excitation temperature, $T_{ex}$. However, the rotational
levels of ammonia are populated according to two temperatures: the
excitation describing the population within a given $K-$ladder and the
rotational temperature connecting the different $K-$ladders. To first
approximation, in the low temperature limit applicable to the
Chamaeleon sources, the correction factor to the partition function
can be computed including only the $K=1$ and 2 ladders,
\begin{equation}
  Q(T_{ex},T_{rot}) = Q(T_{ex})~C(T_{rot}) = Q(T_{ex})~\left( 1 + {g_{2} \over
    g_{1}} \exp(-{{E_{2}-E_{1}} \over kT_{rot}}) \right)~, 
\end{equation}
  
\noindent where the $Q(T_{ex})$ is the standard partition function
from the spectroscopic catalog and indices $1$ and $2$ refer to the
lowest energy metastable levels in the $K=1$ and 2 ladders,
respectively. For $T_{rot} = 11$~K, the correction factor $C = 1.038$,
suggesting that most of the population is within the $K=1$ ladder. The
population in the ground state of the $K=3$ ladder is less than 0.1\%
of that in the ground state of the $K=1$ ladder. \cmark For
consistency with other molecules, the total NH$_3$ column densities
reported in Table~\ref{tab:cden} are computed from opacity corrected
line intensities of the (1,1) line using the Weeds package, and applying
the correction factor $C$, derived above. \umark

\begin{table}
\begin{center}  
\caption{Ammonia column densities in the Chamaeleon cores.} 
\label{tab:ammomms1}
\begin{tabular}{ccc}
\hline \hline 
  \rule[-3mm]{0mm}{8mm}
  Transition  &	$N_l$(MMS1) & $N_l$(C2) \\
	      & (cm$^{-2}$) & (cm$^{-2}$)\\
\hline 
  NH$_3$ (1,1) & $1.41 \times 10^{14}$ & $5.43 \times 10^{13}$ \\
  NH$_3$ (2,2) & $5.39 \times 10^{12}$ & $2.08 \times 10^{12}$ \\
  NH$_3$ (3,3) & $7.41 \times 10^{10}$ & -- \\
\hline
\end{tabular}
\end{center}
Note: Calculations assume $T_{ex}$=7.6~K and 5.5~K for Cha-MMS1 and Cha-C2,
respectively, as derived from the HFS fit to the NH$_3$ (1,1) lines.
\end{table}

\begin{figure*}
\centering

\includegraphics[trim=2.7cm 9.5cm 6cm 3cm, clip=true,
width=0.8\textwidth,angle=0]{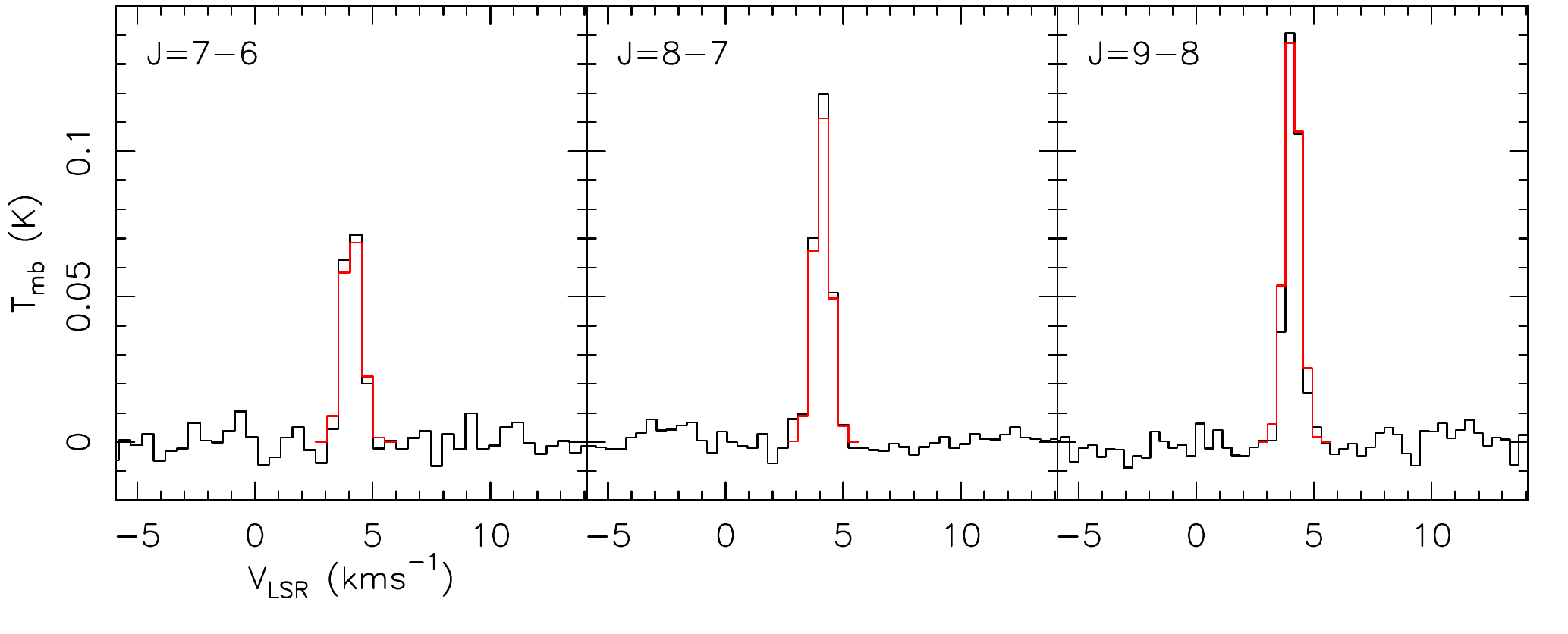}  

\caption{Observed spectra of the J=7--6 to 9--8 lines of HC$_5$N in
  Cha-MMS1 (black histograms) compared to model spectra generated
  using best-fit values from the Python MCMC fit (red histograms).}
\label{fig:hc5n}
\end{figure*}

\section{Line stacking analysis}\label{app:stacking}

We adapted the TMC-1 Markov Chain Monte Carlo fitting and LTE spectral
simulator scripts of Loomis et al. (2021), refactoring the codebase
and improving user accessibility. Originally tailored for Green Bank
Telescope (GBT) observations of TMC-1, the code has been restructured
to support a range of customizable user inputs, making it adaptable to
new sources. It has also been supplemented with documentation. A
user-friendly logging system with real-time progress tracking and
upgraded file management was introduced to handle various MCMC runs
and automate all statistical preprocessing. Walkers have been
reconfigured to consistently initialize within physical bounds,
reducing the need for manual fine-tuning between molecular species. To
address the degeneracy between source size and column density, users
may fix the source size if it can be determined externally, thereby
tightening constraints on column density, or leave it free if unknown.
Column density may also be initialized via maximum likelihood
estimation to further stabilize the inference process. The result is a
robust open-source software tool for MCMC inference of spectra,
successfully validated on the GOTHAM dataset, and publicly available
on GitHub\footnote{https://github.com/KahaanGandhi/Chamaeleon-MCMC}.

To apply the Python-based MCMC tool to the DSN dataset, HC$_5$N was
selected as a benchmark molecule due to the presence of multiple
spectral lines within the DSS-43 frequency range. This choice was
strategic for several reasons: 1) the MCMC algorithm fits all emission
lines simultaneously, diverging from traditional single-line methods;
2) each line possesses a high signal-to-noise ratio; 3) as HC$_5$N is
a member of the cyanopolyyne family, accurate parameter estimation for
this molecule provides informed initial guesses for parameter space
exploration of structurally related, larger molecules like HC$_7$N or
HC$_9$N. 

The MCMC workflow begins with data reduction and preparation,
following the methodology of the GOTHAM survey analysis
\citep{loomis21}. We performed an initial spectral simulation across
the entire DSN bandwidth using the LTE spectral simulator configured
with DSS-43 telescope properties. For each transition exceeding a
threshold intensity, a spectral window of $4.1 \pm 1.5$ km\,s$^{-1}$ was
defined and a local noise level was calculated in the vicinity of the
line.

Following spectral reduction, we applied a Python-based MCMC script to
derive posterior distributions and covariances for the five free
parameters: source size, column density ($N_c$), excitation
temperature ($T_{ex}$), source velocity ($V_{LSR}$), and line width
($\Delta V$). We implemented the affine-invariant ensemble sampler
provided by the emcee toolkit (Foreman-Mackey et al. 2013), exploring
the parameter space with 128 walkers for 10,000 steps. Non-informative
priors were used to maintain physically plausible best-fit values. The
log-likelihood function was defined as the negative half of the sum of
squared residuals between observed and modeled spectra, weighted by
the inverse variance.

For HC$_5$N in both MMS1 and C2 cores, we begin with a “template run”
--- initializing walkers in a wider range using no prior information or
assumptions to prioritize a thorough exploration of the parameter
space. After assessing the quality of these fits visually through
corner plots and ensuring that the parameter distributions align with
our physical expectations, the posteriors from these template runs are
used as priors for subsequent analyses of HC$_5$N and HC$_7$N, and can also
be extended to larger cyanopolyynes. The best-fit values align with
previous surveys, successfully benchmarking the MCMC algorithm and
providing confidence for future applications. The 50th percentile
parameter values for Cha-MMS1 are overlaid with the reduced DSN
spectra in Figure~\ref{fig:hc5n}.

\begin{figure}
\centering

\includegraphics[trim=0cm 0cm 0cm 0cm, clip=true,
width=\columnwidth,angle=0]{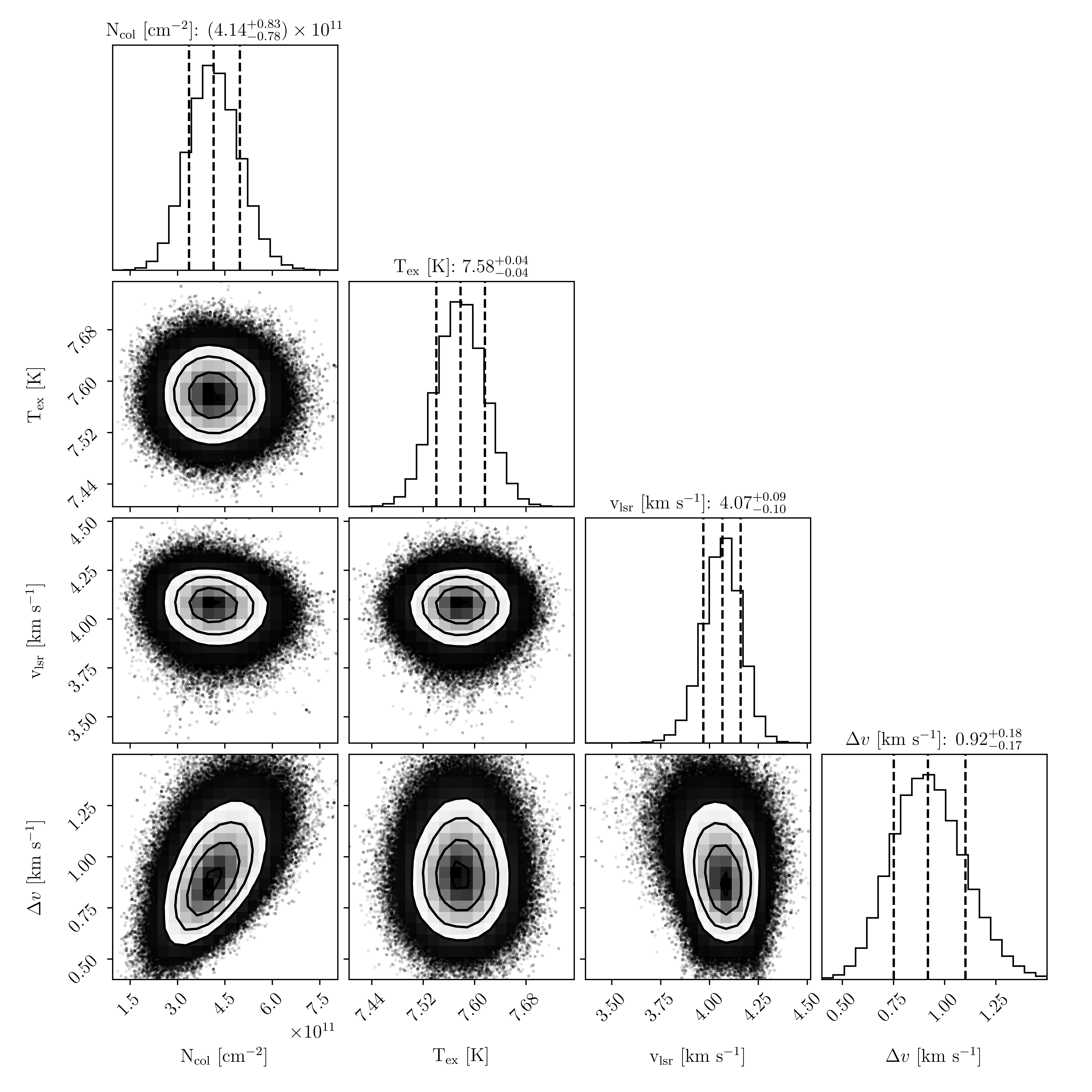}  

\caption{Corner plot for the MCMC fit of HC$_7$N parameter covariances
  and distributions in the Cha-MMS1. The diagonal shows the
  probability distributions of each parameter as histograms, with
  vertical lines marking the 16th, 50th, and 84th percentiles,
  corresponding to $\pm 1 \sigma$ for a Gaussian posterior. The
  off-diagonal scatter plots depict the correlations between pairs of
  parameters, with each axis representing one of the fitted
  parameters. The beam-averaged best-fit column density is consistent
  with the Weeds pencil beam value listed in Table~\ref{tab:cden}.}
\label{fig:corner}
\end{figure}

Having verified that the MCMC approach successfully derives parameter
values in agreement with previous higher-frequency surveys of
Chamaeleon \citep{cordiner2012}, we applied it to the detection of
HC$_7$N. This longer cyanopolyyne presents 8 emission lines in the
DSS-43 frequency range, with 7 falling within the region relatively
free of excess noise. The posteriors from the HC$_5$N template run
serve as the priors for the HC$_7$N analysis. The resulting corner
plot for HC$_7$N in Cha-MMS1 is shown in Fig.~\ref{fig:corner}, with
similar posterior distributions for most parameters compared to
HC$_5$N, and a column density an order of magnitude lower, consistent
with those discussed in Sect.~5.

For this detection, we also employ spectral line stacking to further
boost the signal-to-noise ratio. This technique aligns multiple
transitions along the velocity axis, enhancing the overall signal by
averaging across several lines. In our analysis, we stack the
rotational transitions from $J = 16 - 15$ to $23 - 22$, normalizing
individual spectra to the brightest line and weighting by 1/$\sigma^2$
for the final average, as shown in Fig. 4. This approach leads to a
robust $9.1 \sigma$ detection of HC$_7$N in Cha- MMS1 and a $7.5 \sigma$
detection in Cha-C2 (Fig.~\ref{fig:spemms1} and \ref{fig:spec2},
lower-right panels). The MCMC method shows great
promise for detecting other heavy molecules.

\end{appendix}

\end{document}